\newif\ifapj
\documentclass[preprint]{aastex}

\usepackage[colorlinks,urlcolor=blue,citecolor=black,linkcolor=blue]{hyperref}
\usepackage{amsmath}
\usepackage[export]{adjustbox}
\usepackage{graphicx}

\newcommand{\hi}{H{\sc{i}}}

\newcommand{\water}{\ensuremath{{\rm H_2O}}}
\newcommand{\hm}{\ensuremath{{\rm H}_2}}

\newcommand{\ammonia}{\ensuremath{{\rm NH_3}}}
\newcommand{\cotwo}{\ensuremath{{\rm CO_2}}}
\newcommand{\ctwohtwo}{\ensuremath{{\rm C_2H_2}}}

\newcommand{\ag}{\ensuremath{a_{\rm g}}} 	

\newcommand{\Sd}{\ensuremath{S_{\rm d}}}
\newcommand{\rhod}{\ensuremath{\rho_{\rm d}}}
\newcommand{\rhog}{\ensuremath{\rho_{\rm g}}}

\newcommand{\Lya}{\ensuremath{{\rm Ly}\alpha}}

\newcommand{\ntwo}{\ensuremath{{\rm N}_{2}}}

\newcommand{\LFUV}{\ensuremath{L_{\mathrm{FUV}}}}

\newcommand{\Lx}{\ensuremath{L_{\rm X}}}
\newcommand{\Tx}{\ensuremath{T_{\rm X}}}

\newcommand{\ah}{\ensuremath{\alpha_{\rm h}}}

\newcommand{\um}{\ensuremath{\,\mu}m}

\newcommand{\Msun}{\ensuremath{\,M_{\odot}}}
\newcommand{\MSun}{\ensuremath{\,M_{\odot}}}

\newcommand{\RSun}{\ensuremath{\,R_{\odot}}}
\newcommand{\kms}{\ensuremath{{\,\rm km\,s}^{-1}}}

\newcommand{\ergps}{\ensuremath{{\rm \,erg}\,{\rm s}^{-1}}}

\newcommand{\sqcm}{\ensuremath{{\rm \,cm}^2}}
\newcommand{\psqcm}{\ensuremath{{\rm \,cm}^{-2}}}
\newcommand{\phpsqcm}{\ensuremath{{\rm \,photons\,cm}^{-2}}}

\newcommand{\gpsqcm}{\ensuremath{\rm \,g\,cm^{-2}}}

\newcommand{\pyr}{\ensuremath{{\rm yr}^{-1}}}
\newcommand{\be}{\begin{equation}}
\newcommand{\ee}{\end{equation}}

\shorttitle{FUV AND HEAT SIGNATURE OF ACCRETION IN DISKS} 
\shortauthors{NAJITA \& \'AD\'AMKOVICS}

\begin{document}

\title{FUV Irradiation and  
the Heat Signature of Accretion in Protoplanetary Disk Atmospheres} 

\author{ Joan R.\ Najita$^1$ and 
M\'at\'e \'Ad\'amkovics$^2$ }
\affil{
$^1$National Optical Astronomical Observatory, 
	950 North Cherry Avenue, Tucson, AZ 85719 \\
 	najita@noao.edu \\
$^2$Astronomy Department, 501 Campbell Hall, 
	University of California, Berkeley, CA 94720 \\
	mate@berkeley.edu 
}

\begin{abstract} 
Although stars accrete mass throughout the first few Myr of their lives, 
the physical mechanism that drives disk accretion in the T Tauri phase 
is uncertain, and 
diagnostics that probe the nature of disk accretion have been elusive, 
particularly in the planet formation region of the disk. 
Here we explore whether an accretion process such as the magnetorotational 
instability could be detected through its ``heat signature'', the 
energy it deposits in the disk atmosphere. 
To examine this possibility, we 
investigate the impact of 
accretion-related mechanical heating 
and 
energetic stellar irradiation (FUV and X-rays) 
on the thermal-chemical properties of disk atmospheres at 
planet formation distances. 
We find that stellar FUV irradiation (\Lya\ and continuum), 
through its role in heating and photodissociation,
affects much of the upper warm (400--2000\,K) molecular layer 
of the 
atmosphere, and the properties of the layer are generally in
good agreement with the observed molecular emission features 
of disks at UV, near-infrared, and mid-infrared wavelengths. 
At the same time, the effect of FUV irradiation 
is restricted to the upper molecular layer 
of the disk, even when irradiation by \Lya\ is included. 
The region immediately below the FUV-heated layer  
is potentially dominated by accretion-related mechanical heating.  
As cooler (90-400\,K) CO, water, and other molecules are potential 
diagnostics of the mechanically-heated layer, 
emission line studies of these diagnostics 
might be used to search for evidence of the magnetorotational 
instability in action. 
\end{abstract}


\slugcomment{\today}
			 
\section{INTRODUCTION}

The gas-rich disks that surround young stars at birth play a starring role 
in the build up of stellar masses and in the origin of planets. 
Mass is transported through the disk onto the star, and 
the gas and dust in the disk is available for planet formation.
Emission lines from disk atmospheres offer clues to the 
dynamical state of the gas (e.g., probing disk rotation, 
turbulent line broadening), 
its chemical transformation (e.g., synthesis of organic molecules), 
and the processes that determine the thermal properties of the atmosphere 
(Rab et al.\ 2016; Carmona 2010; Najita et al.\ 2007.)

The thermal structure of the disk affects both its chemistry 
and observational signatures. 
Observations 
reveal that the gaseous atmosphere in the planet formation 
region of the disk reaches higher temperatures than the dust, a result of 
poor gas-grain thermal coupling at the low density conditions 
in the upper disk atmosphere 
(Kamp \& van Zadelhoff 2001; Glassgold \& Najita 2001). 
UV-fluorescent \hm\ emission lines indicate the presence 
of hot \hm\ (1500--3500\,K) within 1\,AU of the star  
(Herczeg et al.\ 2004; Schindhelm et al.\ 2012).  
Near-infrared (2--5\,\um) emission lines from simple molecules 
(CO, OH, water) that arise primarily from within 0.3\,AU also reveal 
that the gaseous atmosphere reaches temperatures of $1500-3000$\,K 
(e.g., Carr et al.\ 1993, 2004; Najita et al.\ 2003; Salyk et al.\ 2009; 
Doppmann et al.\ 2011). 
Similarly, mid-infrared emission from 
water, OH and other molecules (10--20\,\um) within $\sim 1$\,AU 
has a characteristic temperature of $\sim 500-1000$\,K 
(Carr \& Najita 2011; Salyk et al.\ 2011a).  
These temperatures are higher than the dust temperature at the 
same disk radii (e.g., D'Alessio et al.\ 2006). 

Various mechanisms have been proposed to account for the elevated 
gas temperatures, including irradiation by stellar X-ray and 
FUV photons 
(Glassgold et al.\ 2004; 
Kamp \& Dullemond 2004; 
Gorti \& Hollenbach 2008; 
Nomura et al.\ 2007; 
Ercolano et al.\ 2009;
Woitke et al.\ 2009; 
Woods \& Willacy 2009; 
Heinzeller et al.\ 2011; 
Walsh et al.\ 2012; 
Akimkin et al.\ 2013; 
Du \& Bergin 2009) 
The stellar FUV emission from accreting T Tauri stars is powered by 
stellar accretion, whereas their X-ray emission is powered by 
stellar activity. 
The disk can also be heated by 
accretion-related mechanical heating, either from a 
wind blowing over the disk or through the dissipation of gravitational 
energy as material accretes through the disk 
(Glassgold et al.\ 2004).

Historically, thermal-chemical studies of disk atmospheres have focused 
primarily on the role of heating by FUV continuum photons through 
the photoelectric effect on grains and PAHs or \hm\ photodissociation 
(e.g., Kamp \& van Zadelhoff 2001; Jonkheid et al.\ 2004).  
Photoelectric heating is an 
attractive mechanism given the higher luminosity of stellar FUV 
emission 
($L_{\rm FUV}\simeq 10^{31}\ergps$; e.g., Yang et al.\ 2012)
compared to stellar X-rays 
($L_X \simeq 10^{30}\ergps$; e.g., Wolk et al.\ 2005)  
and its natural 
ability to concentrate energy deposition at the top of the atmosphere. 
However, efficient photoelectric heating requires abundant grains in the 
disk atmosphere, in potential conflict with the short settling times of grains 
(e.g., Chiang \& Goldreich 1997; D'Alessio et al.\ 1999).    

The suspicion that disk atmospheres are significantly depleted in 
grains, and the consequent reduction in photoelectric heating efficiency, 
led Glassgold et al.\ (2004) to explore X-ray irradiation and mechanical 
heating as alternative processes.
Significant grain settling in T Tauri disks, 
by factors of $10^{-3}-10^{-2}$ relative to interstellar conditions, 
is indeed inferred observationally (Furlan et al.\ 2005). 
However, the recent recognition that molecules become the primary 
FUV absorbers at low grain abundances, 
and that the resulting photochemical heating is an efficient way to heat 
disk atmospheres (Bethell \& Bergin 2009; Glassgold \& Najita 2015; 
Najita et al.\ 2009) returns FUV to prominence 
in heating the disk surface (\'Ad\'amkovics et al.\ 2014, 2016; 
hereafter AGN14 and ANG16, respectively). 

Nevertheless, the role of mechanical heating in disk atmospheres 
remains of interest, not only 
to understand the thermal structure of disk atmospheres, but also as a 
potential probe of the mechanism that drives disk accretion 
(e.g., Glassgold et al.\ 2004; Bai \& Goodman 2009). 
Despite clear evidence for continued accretion onto stellar 
surfaces throughout the first few Myr of a star's life, 
the physical process that redistributes angular momentum in T Tauri disks 
(and thereby allows accretion to occur)
remains frustratingly unknown (see Turner et al.\ 2014 for a review). 
Thus it is of interest to explore whether the accretion process 
might reveal clues about its nature 
through its energy dissipation in the disk atmosphere. 

While all accretion processes release gravitational energy, the different 
candidate processes differ in where they 
deposit their accretion energy (Turner et al.\ 2014).  
The magnetorotational instability (MRI; Balbus \& Hawley 1992) 
dissipates accretion energy 
higher in the disk, on average, than the height at which accretion occurs, 
a result of the buoyancy of the magnetic field (Hirose \& Turner 2011). 
In addition, the active region of the disk 
at planet formation distances of $\sim 1$--10\,AU 
is restricted to a surface layer, because deeper layers are insufficiently 
ionized to participate in the instability (Gammie 1996). 
As a result, the energy dissipation is further concentrated toward the 
disk surface.

The volumetric heating rate due to the MRI can be quite large, with 
potential observable consequences. 
Hirose \& Turner (2011) estimate a dissipation rate in the 
disk atmosphere equivalent to $\alpha_h \sim 1$, where $\alpha_h$ is 
the accretion-related mechanical heating parameter, defined in 
Glassgold et al.\ (2004) as 
$\Gamma_{\rm acc} = 9/4 \alpha_h \rho c^2 \Omega$ 
where $\rho$ is the local mass density, 
$c$ is the isothermal sound speed, and 
$\Omega$ is the angular rotation speed of the disk. 
In comparison, the value of the viscosity parameter for accretion $\alpha,$ 
when vertically averaged over the column density of the {\it entire} disk, 
is much smaller, $\sim 10^{-2},$ based on T Tauri accretion rates 
(Hartmann et al.\ 1998). 
Recent theoretical studies find that the MRI may be weak or inoperative 
in the planet formation region, creating a potential bottleneck for 
disk accretion, which may be solved by magnetically-driven winds 
(e.g., Bai \& Stone 2013; Kunz \& Lesur 2013; Lesur et al.\ 2014; 
Gressel et al.\ 2015). 
Developing observational probes of 
MRI activity in the planet formation region of the disk may allow us to 
observationally verify or challenge this scenario. 

Here we explore possible observational signatures of mechanical 
heating in disk atmospheres in the context of irradiation heating. 
We make use of our recently 
updated thermal-chemical model of a disk atmosphere 
that is heated by stellar FUV and X-rays  
as well as accretion-related mechanical heating (ANG16). 
We previously used the model to study the impact of stellar \Lya\ irradiation 
on the disk atmosphere at small radii ($\sim 0.3$\,AU), 
implementing the \Lya\ radiative transfer in a simple approach that
includes scattering by \hi\ and absorption by molecules and dust. 

With this approach, we found that the increased heating by \Lya\ 
produced a new component of the model atmosphere close to 
the star: hot (1500--2500\,K) molecular gas (ANG16). 
The properties of the hot component (temperature, column density, 
emitting area) may help to explain the origin of the UV fluorescent 
\hm\ emission that is detected commonly from classical T Tauri stars 
(Herczeg et al.\ 2004; Schindhelm et al.\ 2012).  The observed 
UV \hm\ emission is well explained by strong dipole electronic transitions 
pumped by \Lya\ from excited vibrational levels of \hm.
While the vibrational levels have been assumed to be populated 
non-thermally, by UV and X-ray irradiation (Nomura et al.\ 2007),  
our results suggest that the \hm\ vibrational levels can be 
excited though a thermal pathway. 

Here we use the same model to search for regions of the disk 
atmosphere that are dominated by accretion-related mechanical 
heating and to identify potential observational diagnostics of 
those regions. 
We also compare the warm columns predicted by our model with 
the properties of observational diagnostics of disk atmospheres 
at UV, near-infrared, and mid-infrared wavelengths.  
In the following sections we describe the thermal-chemical model 
(\S 2) and its properties (\S 3) and discuss the 
its ability to account for known observational diagnostics of 
disks and whether accretion-related mechanical heating generates  
a detectable heat signature (\S 4). 
The representative results and figures presented in \S 3 
focus on the results at 0.5\,AU, which are relevant to the 
mid-infrared diagnostics. 
These are supplemented in Appendix A with additional figures and 
discussion that describe the model results at larger radii. 
Related results at smaller radii were discussed previously in ANG16. 
Appendix B describes in greater detail the observed properties of 
disk atmospheres against which our model results are compared.

\section{\label{s:model}THERMAL-CHEMICAL MODEL}

We use a thermal-chemical model of an X-ray and FUV irradiated disk
that was most recently described in ANG16. The model, which builds on
earlier work described in Glassgold et al.\ (2004, 2009) and
\'Ad\'amkovics et al.\ (2011), 
assumes a static disk density and dust temperature 
structure (D'Alessio et al.\ 1999) along with the stellar and disk parameters 
from ANG16, which are 
listed in Table~1. As in ANG16, the adopted dust
properties reflect the assumption that grains have grown and settled quickly
to the midplane, leaving behind a reduced population of small grains in
the disk atmosphere. The grain size parameter $a_g=0.7\micron$ corresponds to a
reduction in the grain surface area by a factor of 20 compared to
interstellar conditions and a dust surface area per hydrogen nucleus
that is $\Sd\approx 8\times10^{-23}\sqcm$.

\begin{deluxetable}{lcl}
\tablecaption{\label{t:std}Reference Model Parameters}
\tablehead{
Parameter & Symbol & Value}  
\startdata 
Stellar mass           & $M_*$    &  0.5 $\MSun$   \\
Stellar radius         & $R_*$    &  2 $\RSun$     \\
Stellar temperature    & $T_*$    &  4000\,K       \\
Disk mass              & $M_D$    &  0.005 $\MSun$ \\
Disk accretion rate    & $\dot M$ &  10$^{-8}\Msun\,\pyr$\\
Dust to gas ratio      & $\rhod / \rhog$  & 0.01 \\
Dust grain size        & $\ag$    &  0.7 \micron   \\
Dust extinction        & $Q_{\rm ext}$ & 1.0 \\
X-ray temperature      & $\Tx$    &  1 keV         \\
X-ray luminosity       & $\Lx$    &  2 $\times 10^{30}\, \ergps$ \\
FUV continuum luminosity\tablenotemark{a} & \LFUV  &  1 $\times 10^{31}\, \ergps$ \\
\Lya/FUV continuum\tablenotemark{b}& $ \eta $  &  3 \\
Accretion heating      & $\ah $   &  0.5
\enddata
\tablenotetext{a}{The FUV continuum luminosity is integrated from 1100--2000\,\AA\ and excludes 
                  \Lya, so that it is smaller than the value used in AGN14.}
\tablenotetext{b}{The ratio of the unattenuated downward \Lya\ photon number flux to 
                  the radially-propogating FUV continuum number flux in 1200-1700\AA\ band.}
\tablenotetext{c}{The total disk column density varies with disk radius as 
$\Sigma \approx 100 (r/{\rm AU})^{-1} \gpsqcm $.}
\end{deluxetable}

With these assumptions, 
we solve the thermal and chemical rate equations to determine the gas
temperature and species abundances. The calculations include the
FUV photochemistry of water and OH (AGN14) as well as additional
abundant molecules and atoms (ANG16). We adopt the photochemical
heating rates detailed in Glassgold \& Najita (2015; hereafter GN15). 
Most notably, as in ANG16, we consider the radiative transfer and
photochemistry of \Lya\ separately from the FUV continuum.

The stellar FUV emission from accreting T Tauri stars has contributions
from both the continuum and the \Lya\ line, although it is the
latter that dominates the luminosity (Herczeg et al.\ 2004). 
As it propagates radially away from the star, the FUV continuum 
is attenuated by dust and molecules in the disk atmosphere. 
The attenuation by an absorber $a$ depends on the 
line-of-sight column of the absorber 
from the star through the atmosphere $N_{a,{\rm los}}$. 
To specify the line-of-sight columns in our model, 
we calculate the ratio of the 
column densities of hydrogen nuclei along the line-of-sight and 
vertically in the D'Alessio atmosphere $N_{\rm H,los}/N_{\rm H}.$ 
The line-of-sight column of the absorber is then approximated 
as $N_a (N_{\rm H,los}/N_{\rm H}),$ where $N_a$ is the vertical 
column density of the absorber.

Studies that reconstruct the stellar \Lya\ emission 
find that it accounts for $\sim 90$\% of the FUV emission from 
accreting T Tauri stars. 
\Lya\ photons leaving the star are scattered in an atomic wind before 
reaching the disk, with a large fraction of the \Lya\ scattered 
away from the disk. 
As a result, the \Lya\ flux is comparable to the flux of FUV 
continuum photons at the transition from atomic to molecular conditions 
in the disk atmosphere (Schindhelm et al.\ 2012; 
Bethell \& Bergin 2011, hereafter BB11). 
Because the scattered \Lya\ photons reaching the disk travel 
more directly downward, their penetration into the disk is 
increased compared to that of FUV continuum photons, 
which propagate at an oblique angle into the disk (BB11). 

Once in the disk atmosphere, the radiative transfer of \Lya\ is complex. 
In contrast to the radiative transfer of the FUV continuum, 
which is governed primarily by absorption due to dust or molecules, 
\Lya\ photons are also scattered by atomic hydrogen, which is abundant 
at the top of the disk atmosphere. 
As a result of the increased scattering, \Lya\ has a
longer path length through the absorptive medium and is efficiently
attenuated in the upper molecular region of the atmosphere
(BB11, ANG16). 

Here, as in ANG16, we adopt a schematic treatment of \Lya\ radiative
transfer (scattering and absorption) based on the detailed work in the
above studies. 
For simplicity, we assume that \Lya\ photons reaching the disk 
propagate directly downward in contrast to the radially propagating 
FUV continuum photons. 
We also set the ratio of the number
flux of 
\Lya\ photons to the number flux of
FUV continuum photons in the 1200--1700\AA\ band 
to be $\eta=3$ at the top of the atmosphere in our reference model. 

Because \Lya\ photons are scattered by \hi\ in the disk atmosphere, 
the photons have a larger effective path length through 
the atmosphere and a greater likelihood of being absorbed by dust 
and molecules.  We treat this effect in a simple, approximate way. 
As described in ANG16, in order to traverse a grid cell of length
$\ell$ that has a scattering mean free path of $\ell_s = 1/n_s\sigma_s,$
where $n_s$ and $\sigma_s$ are the number density and cross section of 
scatterers,  
a \Lya\ photon will take approximately $\tau_s^2$ steps, where 
$\tau_s = n_s \sigma_s \ell = \ell/\ell_s.$ 
As a result, the photon travels a distance of
approximately $\ell_{\rm eff} = \tau_s^2 \ell_s = \tau_s\ell$ through the
absorbing medium. Because of the longer path length, the effective
optical depth for absorption is approximately
\be 
\tau_{\rm eff} \equiv n_{\rm a} \sigma_{\rm
a} \ell_{\rm eff} = \tau_{\rm a} \tau_{\rm s}, 
\ee 
where $n_{\rm a}$ and $\sigma_{\rm a}$ are the number density and 
absorption cross of absorbers, 
and $\tau_{\rm a}$ is the absorption optical depth. 
The increased path length leads to increased \Lya\ (intensity and) 
absorption, which both attenuates
the \Lya\ and enhances its photochemical and photoelectric heating.

In our earlier study 
focusing on the impact of \Lya\ irradiation at small disk radii,
we found that the increased heating in that region of the disk
produced a new component of the disk atmosphere: hot (1500--2500\,K)
molecular gas that may account for the UV fluorescent \hm\ emission
that is detected commonly from classical T Tauri stars. Because the 
treatment of \Lya\ scattering is approximate, our results should be
considered illustrative rather than quantitative. An improved
treatment that includes scattering effects in a more realistic way
is needed to understand the effect of \Lya\ on the detailed properties
of the atmosphere.  

Our model also includes self-shielding in the 900-1000\AA\ band. 
The opacity in the band is dominated by \hm, CO, \ntwo, and C, 
with \hm, CO, and \ntwo\ self- and mutually shielding one another 
as well as other species. To include these processes, we use the 
\hm\ shielding function from Draine \& Bertoldi (1996), tabulations 
of CO shielding from Visser et al.\ (2009), and \ntwo\ shielding from 
Li et al.\ (2013), following the prescription described in ANG16. 

Water vapor, an important source of FUV opacity, can freeze out 
onto cold grains at dust temperatures of 100--125\,K 
(e.g., Woods \& Willacy 2009; Woitke et al.\ 2009; Meijerink et al.\ 2009; 
Heinzeller et al.\ 2011)
At radial distances within a few AU of the star, the grains in the 
disk atmosphere are warm enough that an insignificant amount 
of water is condensed onto their surface, while
at larger radial distances freezeout becomes more of a concern. We
consider the freezeout of water onto grain surfaces by calculating
the rate of adsorption and balance it with the rate of water removal  
from grains via thermal- and photo-desorption (Fraser et al.\ 2001; 
\"Oberg et al.\ 2009). 

The rate of water adsorption onto grains is determined by the rate
at which molecules strike grains, 
$R_{\rm ads} = n(\water) \bar{v}(\water) S_d$,
where 
$\bar{v}(\water) = \sqrt{ 8kT/18\,\pi\,m_H}$ 
is the mean thermal
velocity of the water molecules and $n(\water)$ is their number density.
We assume unit sticking probability once a water molecule strikes
a grain. Thermal desorption is included as a zeroth order process
based on the experimental results of Fraser et al.\ (2001) and
is implemented with a treatment similar to that of Visser et al.\ (2011).
Photodesorption is determined by the local FUV photon number flux
and a temperature-dependent photodesorption efficiency given
by \"Oberg et al.\ (2009). Water molecules desorb intact 70\% of the
time and otherwise as OH + H.

In addition to the approximate \Lya\ radiative transfer described 
above, there are several other shortcomings of the model. 
One shortcoming is that hydrostatic equilibrium
is not enforced, i.e., when we calculate the gas temperature, we
do not adjust the assumed density structure, that of the D'Alessio
et al.\ (1999) model, to maintain pressure balance.  Because the
gas temperature exceeds the dust temperature in the upper atmosphere, 
we overestimate the density there. We also
assume that the dust abundance and size distribution are the same
at all heights and radii in the atmosphere. Transport of material
(radial or vertical) is not considered, and we assume that the
chemical timescales are rapid enough that abundances and thermal
rates are determined by local conditions.

To explore the impact of not enforcing hydrostatic equilibrium, we
carried out a simple experiment in which we adjusted the density 
profile of the atmosphere to approximate hydrostatic equilibrium
and then recalculated its thermal-chemical structure. We found that
the lack of explicit hydrostatic equilibrium did not greatly 
affect the properties (temperature, abundances) of the atmosphere. 
The recalculated molecular abundances and temperatures 
were within a factor of 2 and within 10\%, respectively, 
of their original values
in the region where the molecules produce most of their emission. 
We estimate that the errors in temperature and abundances that 
result from the lack of hydrostatic equilibrium do not exceed 
those derived from uncertainties 
in other model properties such as the extent of grain growth, 
and the luminosity and propagation of the scattered \Lya\ component 
of the UV field.

\begin{figure}[htb!]\begin{center}
\includegraphics[height=5.25in]{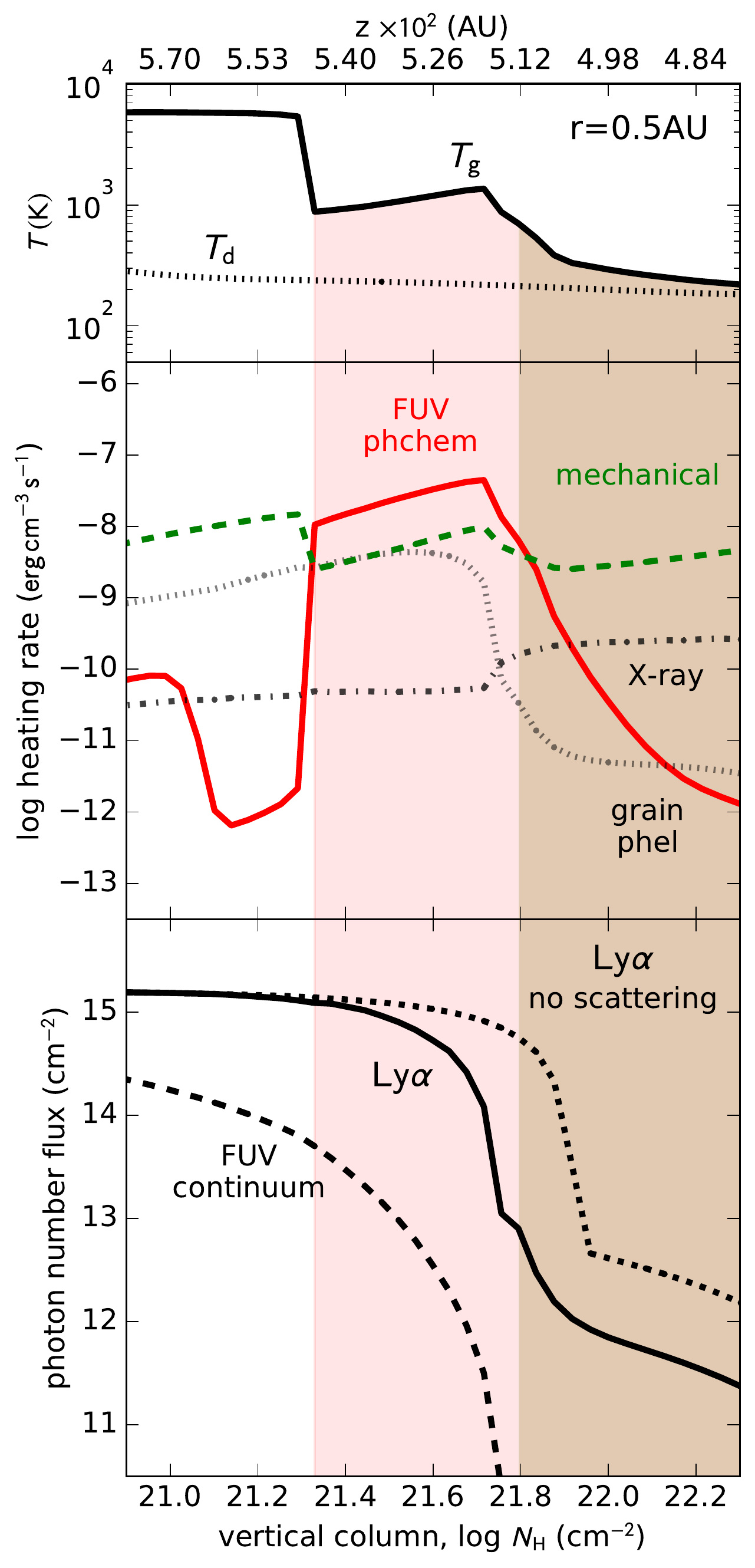}

\caption{\footnotesize Vertical profiles of gas and dust temperatures 
(top panel, solid and dotted lines, respectively), 
heating rates (middle panel), 
and FUV radiation fields (bottom panel) at $r=0.5$\,AU in our 
reference model. 
FUV photochemical heating (red solid line) 
dominates in the warm upper molecular layer of the disk (strawberry shading) 
where the FUV continuum and \Lya\ are absorbed.
Mechanical heating (green dashed line) 
dominates above and below this region. 
X-ray (dash-dotted line) and grain photoelectric heating (dotted line) 
are shown for comparison.
The FUV continuum and \Lya\ are strongly attenuated in the 
photochemically-heated layer.
}
\end{center} \end{figure}

\begin{figure}[htb!]\begin{center}
\includegraphics[height=5.25in]{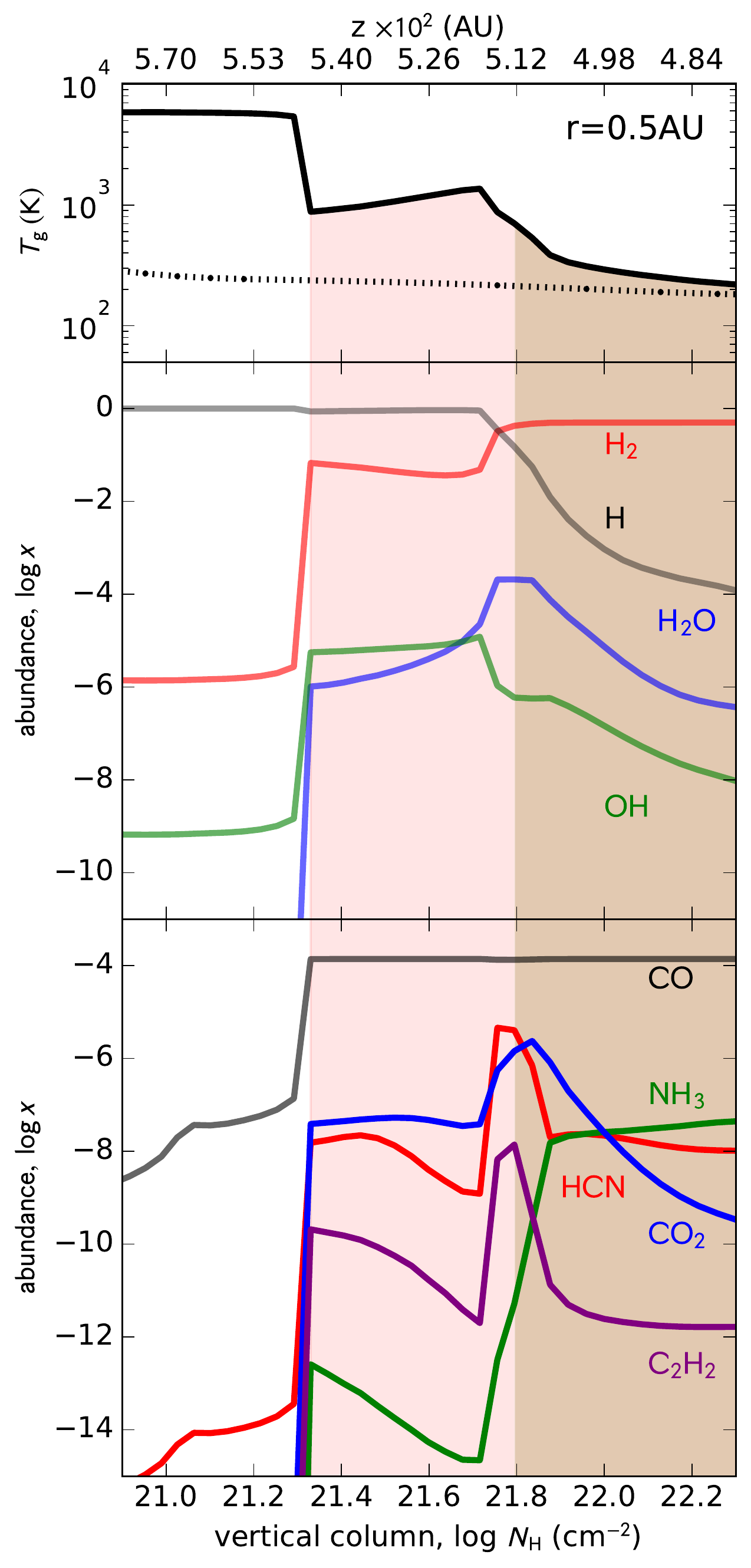}

\caption{\footnotesize Vertical profiles of temperatures (top panel) and abundances 
(middle and bottom panels) at $r=0.5$\,AU in our reference model.  
The warm upper layer where FUV heating dominates (strawberry shading) 
is rich in molecular species that are commonly detected 
from T Tauri disks.
The cooler mechanically heated layer (chocolate shading) lies 
below. 
}
\end{center} \end{figure}

\section{Results}

At high altitude, the gaseous disk atmosphere is hot ($\sim 5000$\,K) 
and in atomic form (Figures 1, 2, top panels, solid lines).
In our reference model, in which $\alpha_h = 0.5$, 
mechanical heating is the primary heating mechanism at the 
top of the atmosphere, although FUV heating is also important 
(Figure 1, middle panel). 
Because molecules are absent in this region of the atmosphere, 
photochemical heating is unimportant, and the FUV heats through 
its interaction with grains via the photoelectric effect. 

At intermediate altitudes (strawberry-colored regions),  
simple molecules (e.g., \hm, OH, \water) are abundant, 
although much of the hydrogen is in atomic form (Figure 2, 
middle panel).  
\Lya\ photons are scattered by the HI in this layer, 
and the OH and \water\ absorb FUV photons (both \Lya\ and continuum; 
Figure 1, bottom panel). 
The resulting photochemical heating dominates the heating 
(Figure 1, middle panel),  
and the atmosphere is very warm as a result ($\sim 800-1000$\,K) 
from 0.5--2\,AU (Figure 1, top panel and Appendix A, Figure 7). 
As they deposit their energy, the 
\Lya\ flux is significantly attenuated through the layer 
and the FUV continuum plummets (Figure 1, bottom panel). 

At lower altitudes,  
immediately below the FUV-heated layer 
(chocolate-shaded region), 
mechanical heating dominates (Figure 1, middle panel), 
and \hm\ is the dominant form of hydrogen (Figure 2, middle panel). 
Lacking significant FUV heating and subject to stronger 
dust-gas collisional cooling because of the higher density, 
the gas in this region of the atmosphere is cooler than 
in the intermediate layer but nevertheless 
warmer than the dust in the same region of the atmosphere 
(Figure 1, top panel, dotted line). 
While \Lya\ is greatly attenuated through the 
FUV-heated layer, the photon flux that survives 
into the layer below is significant ($\sim 10^{12}\phpsqcm$), 
and \Lya\ dominates the FUV field at depth 
(Figure 1, bottom panel; BB11). 
Water declines in abundance in the chocolate-shaded layer 
as a consequence of 
continued photodissociation 
and the declining gas temperature; 
the latter  reduces the neutral formation rate of water 
(see also AGN14). 

\begin{figure}[t]\begin{center}
\includegraphics[height=3.7in]{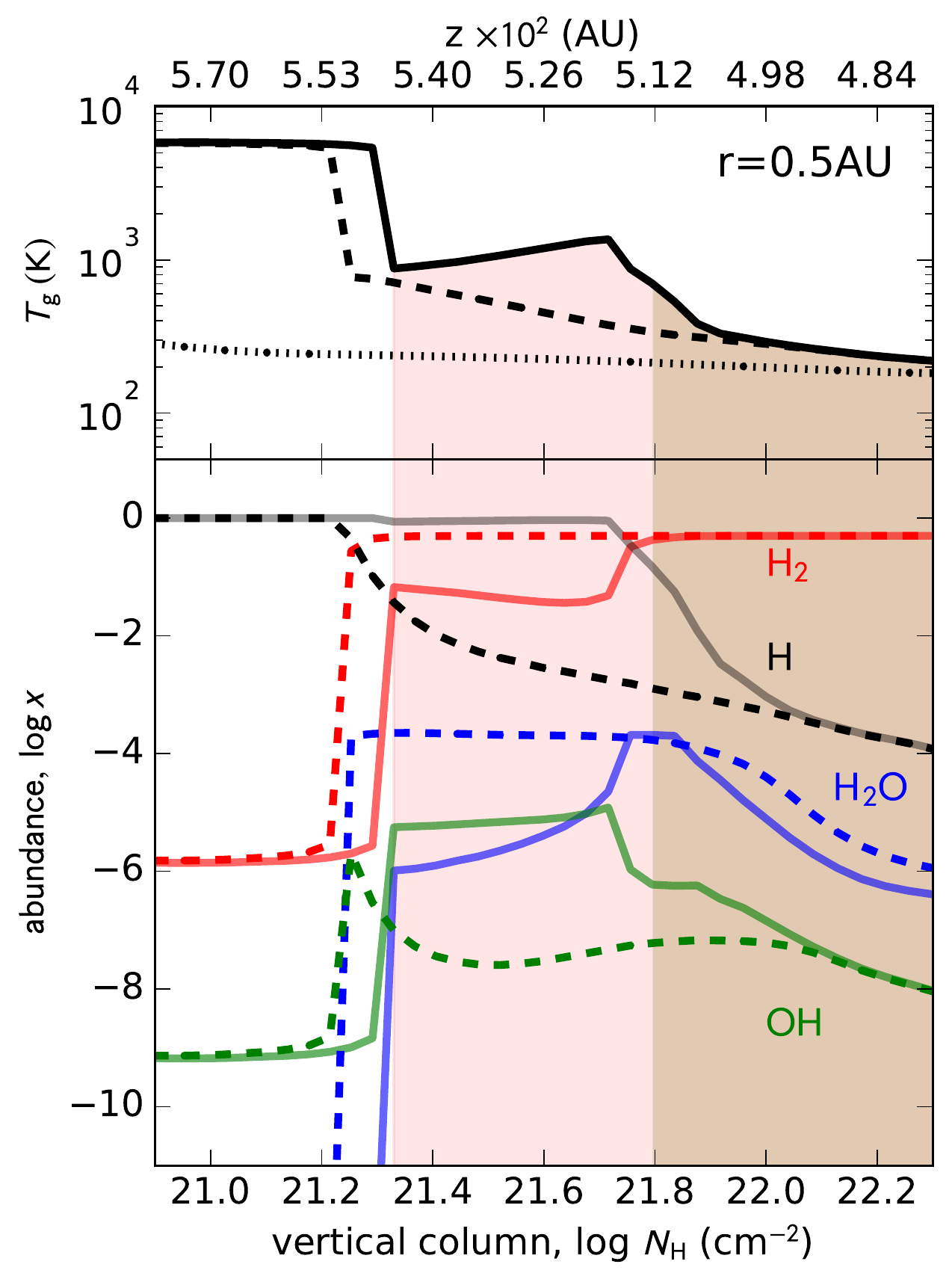}

\caption{\footnotesize Vertical profiles of gas temperatures (top panel) and 
molecular abundances (bottom panel) 
with and without \Lya\ irradiation 
(solid and dashed lines respectively) 
at $r=0.5$\,AU. 
When \Lya\ is not present, the molecular portion of the 
atmosphere is cooler, 
and the molecular transition occurs higher in the atmosphere. 
When present, \Lya\ dissociates \water\ and enhances the 
OH abundance in both the FUV-heated (strawberry shading) and 
mechanically-heated (chocolate shading) layers. 
}
\end{center} \end{figure}

Figure 3 shows the impact of \Lya\ irradiation on the 
thermal-chemical properties of the atmosphere. 
If the \Lya\ component of the FUV field 
were not present, the molecular portion of the atmosphere 
would be cooler (top panel, dashed line). 
Water would be more abundant and atomic H less abundant 
in both the strawberry and chocolate layers at 0.5\,AU to 2\,AU  
(bottom panel, blue and black dashed lines, respectively; 
see also Appendix A, Figure 9). 
OH, a dissociation product of water, would also be much less 
abundant at all radii (bottom panel, green dashed line). 

\begin{figure}[t]\begin{center}
\includegraphics[height=3.7in]{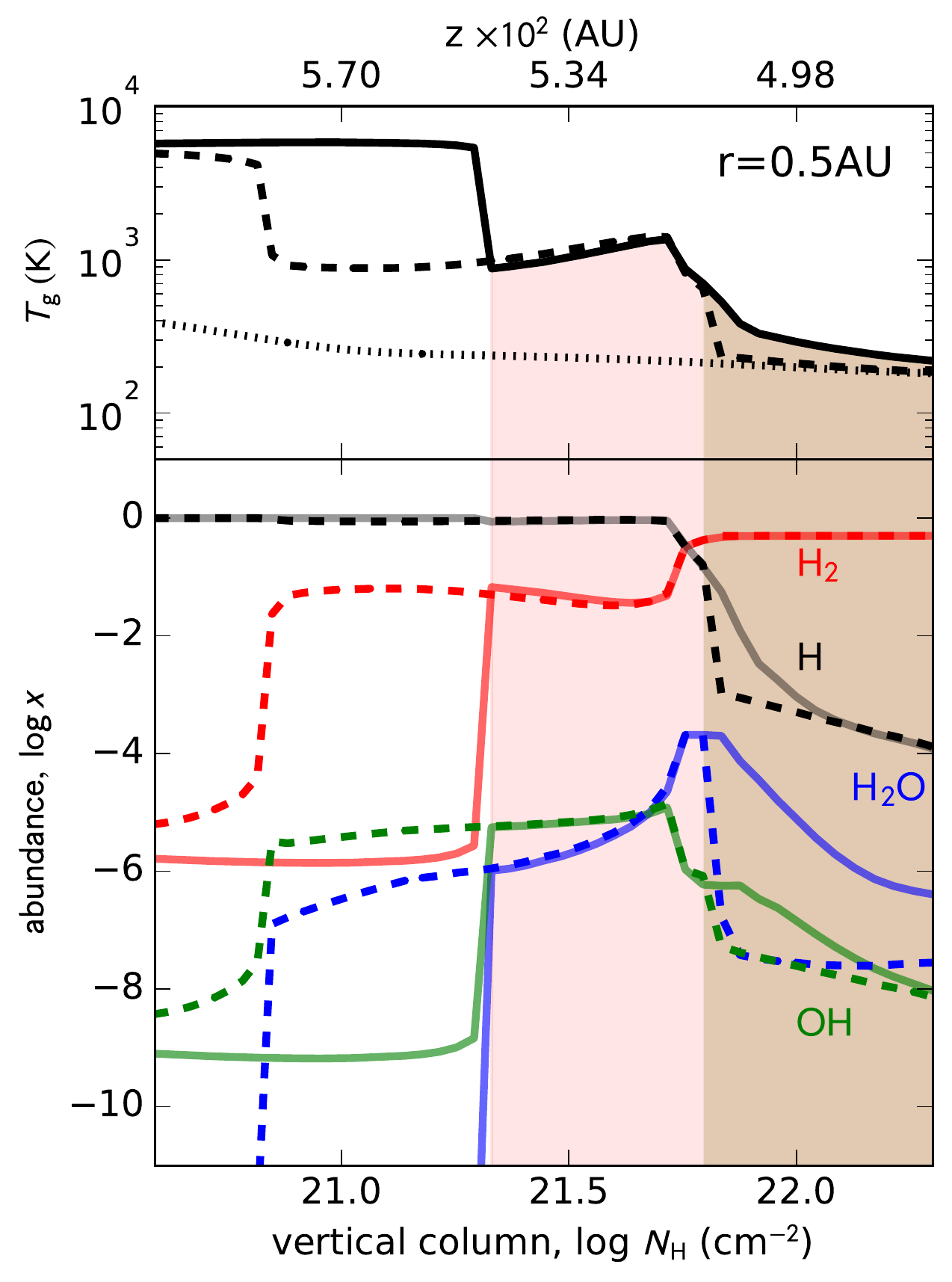}

\caption{\footnotesize Vertical profiles of gas temperatures (top) and 
molecular abundances (bottom) 
with and without mechanical heating 
(solid and dashed lines, respectively) 
at $r=0.5$\,AU. 
When mechanical heating is reduced, the molecular transition 
occurs higher in the atmosphere and the temperature 
is reduced in the molecular region below the 
FUV-heated layer. 
}
\end{center} \end{figure}

Figure 4 shows the impact of mechanical heating on the atmosphere. 
Mechanical heating affects the structure above and below the 
FUV-heated layer (Fig.~1).  At high altitudes,
the transition from the hot atomic layer to the 
FUV-heated layer occurs higher in the atmosphere 
when mechanical heating is absent (dashed line)  
than when it is present (solid line). 

The reason for this difference is that the 
transition from atomic to molecular conditions is regulated by
the balance between \hm\ formation on grains and its
collisional destruction by atomic hydrogen 
($\hm + {\rm H} \rightarrow 3\,{\rm H})$.
The destruction pathway is critically sensitive to temperature, with
the rate coefficient declining by 15 orders of magnitude 
between 3000 and 1000\,K (Palla et al.\ 1983). 
The feedback between increased
molecular cooling and decreased \hm\ destruction leads to the
sharp transition from atomic to molecular conditions (AGN14). 
If mechanical heating increases 
in the region of the transition from atomic to molecular conditions,
\hm\ destruction increases rapidly and the layer becomes 
atomic, pushing the transition deeper towards the midplane. 
Conversely, when mechanical heating is reduced,
the tipping point of the transition is reached higher in the
atmosphere.

\begin{figure}[t]\begin{center}
\includegraphics[height=3.2in]{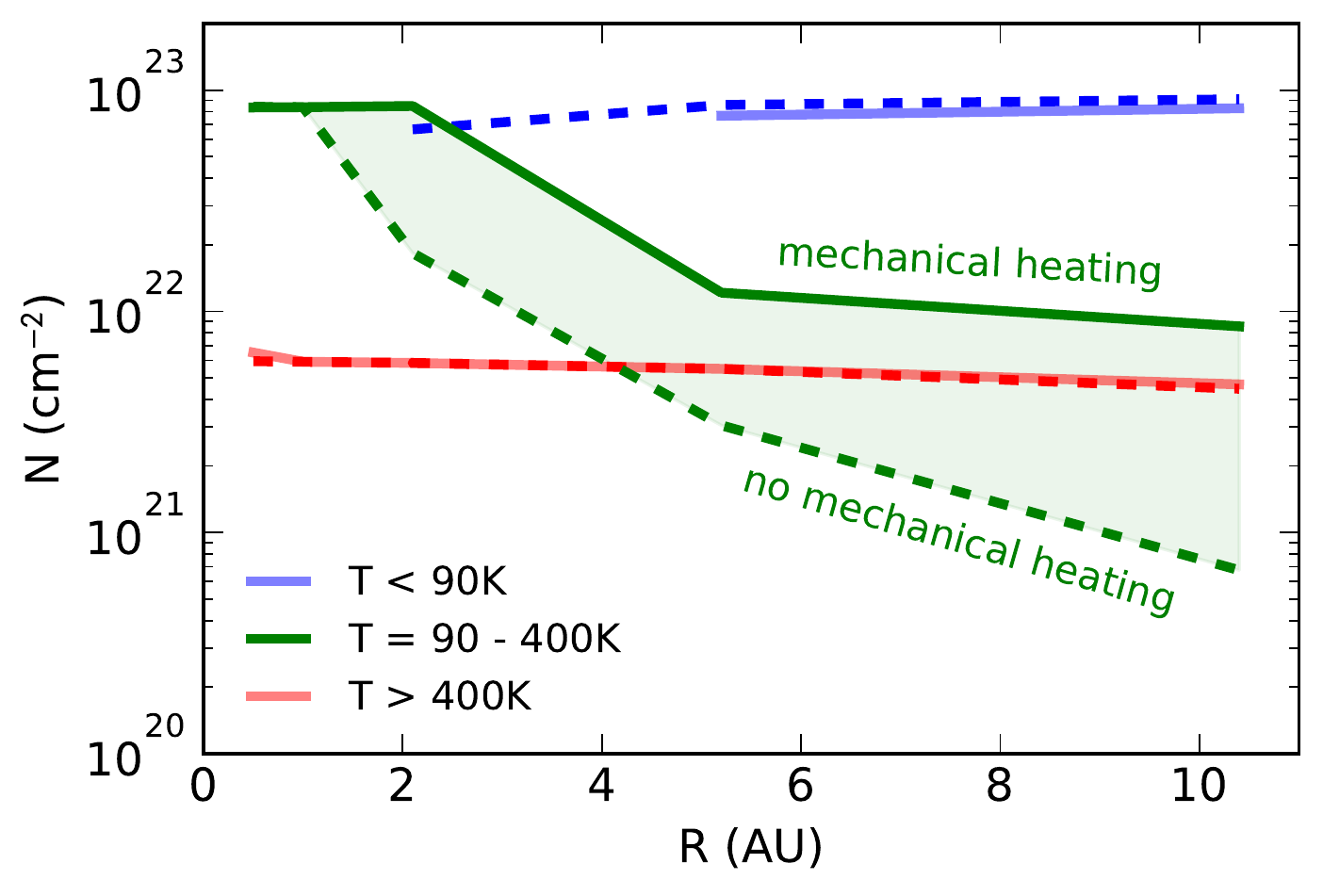}

\caption{\footnotesize Gas column densities in the disk atmosphere 
in several temperature ranges as a function of disk radius
with (solid lines) and without (dashed lines) 
mechanical heating. Mechanical heating enhances the column 
density in the 90--400\,K temperature range. 
}
\end{center} \end{figure}

Deeper in the disk, below the FUV-heated layer, 
the gas temperature at all radii (0.5\,AU to 10\,AU) 
is cooler without mechanical heating 
than when mechanical heating is present. 
As a result, the column density of warm gas 
is enhanced when mechanical heating is present. 
Figure 5 shows that the gas column density in the disk 
atmosphere in the temperature range 90--400\,K is 
large, $\gtrsim 10^{22}\psqcm,$ and 
4--10 times larger beyond 1\,AU when 
mechanical heating is present (solid green line) 
than when it is absent (dashed green line). 
At the smaller radii ($\lesssim 1$\,AU), mechanical heating 
enhances the column density of gas at 
higher temperatures. 
At 0.5\,AU the column density of gas in the 
200-400\,K region is 4 times larger with 
mechanical heating than without. 
To exclude any warm molecular gas that
is present deep in the atmosphere, near the (unobservable) disk
midplane, the column densities shown in the figure are restricted
to the surface $N_H = 10^{23}\psqcm$ of the 
disk. 

CO is an attractive potential diagnostic of the mechanically heated layer.  
Relatively insensitive to other factors 
(e.g., the intensity of \Lya\ flux in the layer, 
or the location of the snow line), 
CO is reliably present in the molecular atmosphere at high abundance 
in and below the FUV-heated layer 
($x_{\rm CO} \sim 10^{-4}$; Figure 2; see also Appendix A, Figure 8). 
At 2--10\,AU, 
the CO column density in the temperature range 90--400\,K 
is $\gtrsim 10^{18}\psqcm$ when mechanical heating 
is present and $\sim 4-10$ times lower when 
it is absent. 

Other potential diagnostics of the mechanically heated layer depend 
more sensitively on other factors. 
For example, the abundance of water in the layer 
depends on both the strength of mechanical heating and the 
flux of \Lya. 
Mechanical heating raises the gas temperature, an effect that 
tends to increase the abundance of water, which  
depends sensitively on temperature (e.g., Glassgold et al.\ 2009). 
Conversely, \Lya\ photodissociates water and reduces  
its abundance. 

As a result of these two effects, the abundance of 
water in the mechanically heated layer varies with 
radius. 
At 0.5\,AU, the \Lya\ flux is reduced significantly, by $\sim 1000$,  
through the FUV-heated layer and the 
mechanically heated layer is warmed to $>250$\,K. 
Because the neutral synthesis pathway for water is active in 
this temperature range, the layer contains a significant column of 
warm water.
At larger radii, the \Lya\ flux is reduced by a smaller factor 
through the FUV-heated layer (by $\sim 100$ at 1\,AU; 
Appendix A, Figure 7), 
the mechanically-heated layer is cooler than 250\,K, 
and little water is present in that layer 
as a result (Appendix A, Figure 9).

\begin{figure}[t]\begin{center}
\includegraphics[height=3.1in]{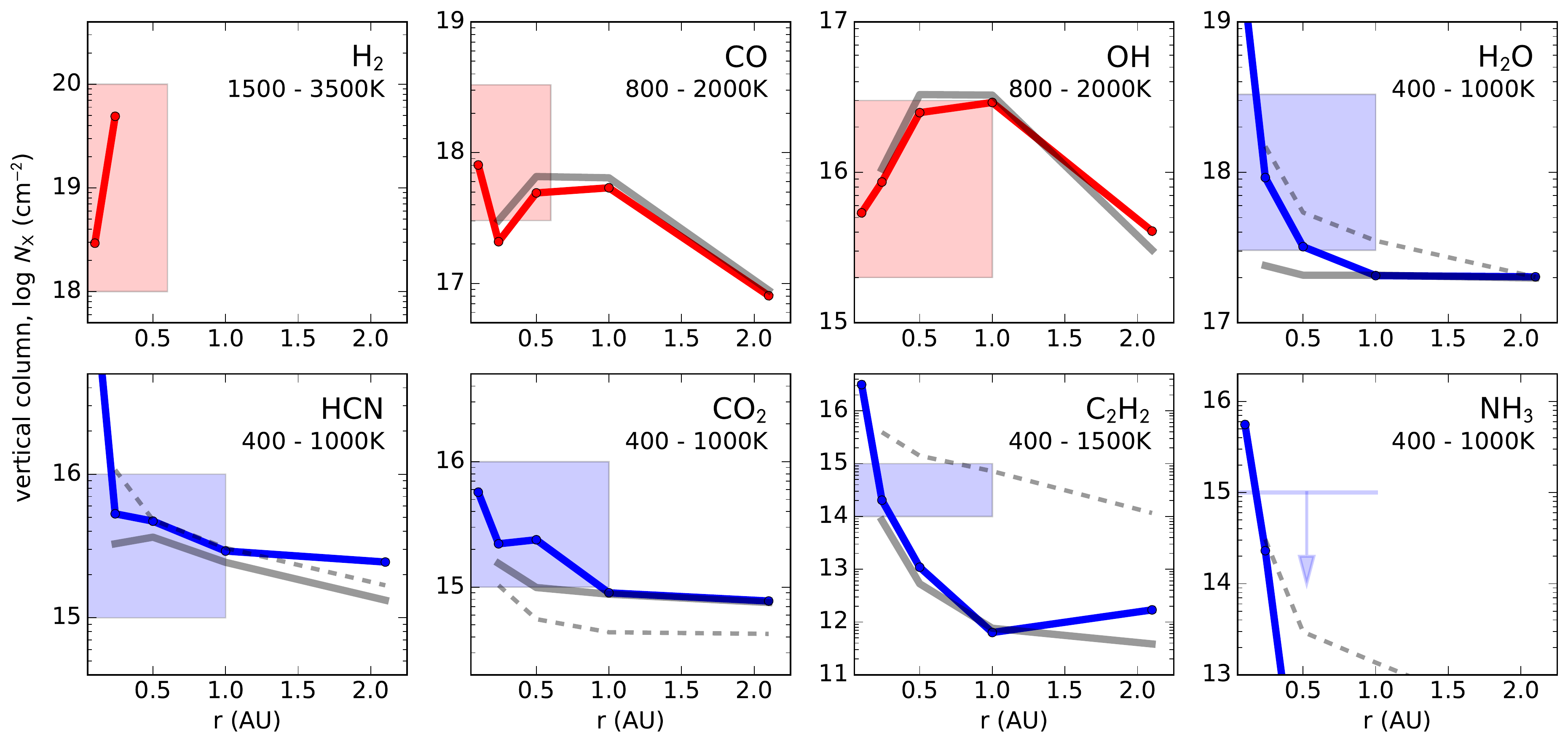}
\caption{\footnotesize Warm gas column densities in the reference model 
within the temperature range inferred for the observed diagnostics 
as a function of disk radius (colored solid lines) compared with 
the range of column densities and emitting radii inferred observationally 
for the diagnostic (shaded region) in the same temperature range. 
The shaded region extends out to the inferred 
value of $R_e$ (see text for details). 
The observational upper limit on the column density of warm \ammonia\ 
is indicated by the downward arrow.  
The model results without mechanical heating 
($\alpha_h = 0.01$; solid gray lines) and 
without \Lya\ (dashed gray lines) are shown for comparison. 
The reference model is in good agreement 
with the observations for most species. 
}
\end{center} \end{figure}

The upper molecular region of the reference model 
(strawberry-shaded region in the figures) captures 
the general properties of the molecular emission 
observed from the inner regions of T Tauri disks. 
Figure 6 compares the warm molecular column densities 
in the disk atmosphere model with 
the warm molecular emission columns inferred from spectroscopy. 
The shaded regions indicate the general properties of 
the molecular emission detected by {\it Spitzer} 
from T Tauri disks 
(OH, \water, HCN, \cotwo, \ctwohtwo, and upper limits 
on \ammonia). The properties are derived from simple 
slab fits to the emission, parameterized by molecular 
column density, emission temperature, and emitting area 
$\pi R_e^2$ (Carr \& Najita 2011, 2008; Salyk et al.\ 2011a). 

In Figure 6, the height of the shaded region indicates 
the typical range of observed molecular emission columns, 
and the shaded region extends out to typical values of $R_e.$
The heavy solid lines show, as a function of disk radius,  
the warm columns from the reference model 
within the temperature range reported for the emission. 
The relevant temperature range is indicated in the legend 
in each panel. 

Figure 6 also shows the results described previously 
for \hm\ and CO (ANG16) using the same approach as for the 
{\it Spitzer} diagnostics described above. 
The large columns of hot \hm\ 
($\log N_{\hm} \sim 19$  at 1500--3500\,K) 
produced within 0.3\,AU in the model 
are comparable to the properties inferred for the hot \hm\ that 
produces the UV fluorescent \hm\ emission from T Tauri stars. 
\hm\ temperatures and column densities of 
$T_{\hm}=2500\pm 1000$\,K and $\log N_{\hm}=19\pm 1$
are inferred from the analysis of UV \hm\ emission 
from T Tauri stars
(Schindhelm et al.\ 2012; Herczeg et al.\ 2004).
The resolved UV \hm\ line profiles of 
normal (non-transition) T Tauri stars 
(FWHM $\sim 50$--70\kms; France et al.\ 2012)
suggest that most of the emission arises from 
within $\sim 0.7$\,AU. 

The CO rovibrational emission from T Tauri stars has line widths  
comparable to those of the UV \hm\ emission, 
indicating that it also arises from primarily within 0.7\,AU.  
Typical (deprojected) CO emission columns and temperatures are
$\log N_{\rm CO} = 17.5-18.5$ and 900--1600\,K, respectively  
(Salyk et al.\ 2011a,b; Najita et al.\ 2003; Banzatti \& Pontoppidan 2015). 
These properties 
are similar to the hot columns of CO found in the reference model. 
In general, the molecular properties of the reference model are 
in good agreement with those of the observed molecular emission 
from T Tauri disks at UV, NIR, and MIR wavelengths. 

The relative roles of FUV and mechanical heating in the reference 
model are also illustrated in Figure 6. 
The results for models with no mechanical heating ($\alpha_h =0.01$) 
and no \Lya\ are shown as light solid and light dashed lines, 
respectively. 
Comparing these results with those of the reference model
shows that the column densities of hot \hm, CO, and warm OH 
are similar whether mechanical heating is present (heavy solid line) 
or not (light solid line), 
i.e., these quantities are dominated by FUV heating. 
There is no dashed line in these panels because mechanical 
heating alone (in the absence of FUV) does not create a warm 
enough atmosphere to contribute molecular emission in the temperature 
ranges indicated. 

Mechanical heating does help to enhance the column density 
of water in the 400--1000\,K temperature range, whereas FUV heating 
alone falls short of the observed warm water column density, 
in part because the FUV dissociates water. 
The column density of warm \cotwo\ is also best matched 
in the model with both mechanical heating and \Lya. 
Although \ctwohtwo\ is underproduced in the 
reference model, it is fairly abundant when \Lya\ is weak. 
Perhaps a model with more modest \Lya\ irradiation beyond 
the hot \hm\ emitting region may 
be able to produce hot \hm\ and CO close to the star, while also  
retaining a reasonable column density of \ctwohtwo\ 
at larger radii. Models with detailed \Lya\ radiative transfer   
would be valuable in exploring this possibility.

\section{Discussion}

We find that stellar FUV irradiation, 
through its role in heating and photodissociation,
affects much of the disk atmosphere that produces 
the well-known molecular emission diagnostics of inner disks.  
In particular, \Lya\ photons, with their distinctive radiative transfer, 
have a significant impact on the thermal-chemical structure of 
the upper atmosphere. 
In ANG16 we found that our simple prescription for \Lya\ 
radiative transfer, which includes scattering by HI and absorption 
by molecules and dust, can account for the large 
column densities of 
hot UV fluorescent \hm\ and rovibrational CO emission 
detected from T Tauri disks at small radii ($\lesssim 0.5$\,AU), 
properties of long standing that have been otherwise difficult 
to explain. 
Here we find that the FUV-irradiated layer of the atmosphere 
at larger radii (out to $\sim 1$\,AU) can also account for the 
{\it Spitzer} OH, \water, HCN, and \cotwo\ emission that is 
commonly detected from T Tauri disks (Fig.~6). 
As described in \S 3, 
while the observations are best fit with a model that includes both 
UV irradiation (\Lya\ and continuum) and mechanical heating, 
the UV (and \Lya\ in particular) does much of the heavy lifting.  

\subsection{OH as a Probe of \Lya\ Irradiation} 

\Lya\ irradiation also produces abundant OH in the planet formation 
region, with the OH/\water\ column density ratio in the atmosphere 
in good agreement with observations. 
Because the upper disk atmosphere has a significant atomic component, 
strong scattering by HI at the disk surface enhances 
\Lya\ absorption by molecules and dust. 
The \Lya\ component deposits energy deeper in the atmosphere than 
the FUV continuum (Figure 1) enhancing the 
temperature of the upper molecular layer 
which rises to $\gtrsim 1000$\,K (Figure 2). 
OH is abundant throughout this region, because 
the FUV energy is deposited through dissociation of water, producing OH. 
FUV irradiation thus produces a layered atmosphere of very warm 
($> 1000$\,K) OH  
with a column density of $\sim 3\times 10^{16}\psqcm$ at 0.5\,AU, 
overlying cooler (1000-500\,K) water, with a column 
density $\sim 5\times 10^{17}\psqcm.$
Thus, when \Lya\ scattering is included, the upper atmosphere 
is hotter and OH is more abundant than when scattering is ignored. 

These results differ from those of AGN14, in which we treated the 
radiative transfer of \Lya\ in the same fashion as the 
FUV continuum, i.e., as pure attenuation. 
In that study, we found a 
shallower photochemically heated layer. 
As a result, the warm column ratio of OH/\water\ was 
$\sim$\,1/300 at $\sim 1$\,AU rather than the ratio of 
$\sim 1/10$ found here 
for the same total stellar (\Lya\ + continuum) FUV luminosity.  
Column density ratios in this range are similar to the 
OH/water emission column density ratio inferred from 
{\it Spitzer} spectra of T Tauri stars 
(Salyk et al.\ 2011a; Carr \& Najita 2008). 
{\it Spitzer} observations indicate an OH emission column density 
$\sim 0.01-0.1$ that of water, assuming that the two diagnostics have 
similar emitting areas, with the OH having a higher average 
temperature than the water as well, as in the model results. 

One limitation of our model is that the \Lya\ radiative 
transfer is treated very approximately. 
For example, we assume that at all disk radii
the number flux of downward-propagating \Lya\ photons 
at the top of the atmosphere is a constant value of 
$\eta=3$ times the number flux of FUV continuum photons, 
following the results of BB11. 
The BB11 calculations are themselves approximate. While  
they include \hi\ scattering and dust absorption 
of \Lya, they do not treat the synthesis of 
FUV-absorbing molecules such as OH and \water. 
The latter shortcoming can be addressed with calculations that integrate
sophisticated \Lya\ radiative transfer in a disk atmosphere 
(e.g., BB11) with detailed chemistry (e.g., as in Du \& Bergin 2014).  

Observations can also help to constrain the role of \Lya\ 
in heating disk atmospheres. Because \Lya\ has such a strong 
impact on the OH column density, high resolution spectroscopy 
of the OH emission detected by {\it Spitzer} can measure the 
OH column density and temperature as a function of radius, 
for comparison with the results obtained here. 
The comparison will lend insight into whether
the treatment of \Lya\ employed here is approximately correct 
and where it fails. 

\subsection{Molecular Probes of Mechanical Heating} 

While stellar FUV irradiation has a significant impact on  
the upper disk atmosphere, its thermal effect is restricted to the  
disk surface even when irradiation by \Lya\ is included. 
With the impact of FUV irradiation thus restricted, 
mechanical heating has the opportunity to imprint 
a signature on the thermal-chemical properties of the layer below it.  
Detailed attention to the properties of this region of the 
atmosphere may open opportunities to detect its effects. 

Our results suggest CO as a potential 
diagnostic of this region. Because its abundance is relatively 
insensitive to UV irradiation and the thermal structure of the 
atmosphere, the column density of warm CO may serve 
as a signature of an atmosphere warmed by mechanical heating. 
At disk radii 1--10\,AU, the CO column density in the temperature 
range 90--400\,K is $\gtrsim 10^{18}\psqcm$ and 
4--10 times higher when mechanical heating is present than when it 
is absent (\S 3). 
CO in the 100--500\,K range is detectable in 
the FIR and submillimeter. 
For example, Meeus et al.\ (2013)
detected CO emission in this temperature range 
from T Tauri stars and Herbig AeBe stars using 
{\it Herschel} PACS. 
Among the T Tauri stars, the observations were only 
sensitive to bright CO emission from energetic sources 
with evidence for winds and outflows (e.g., DG Tau, AS205A). 
More sensitive measurements are needed to detect warm 
CO emission from less active, more typical T Tauri stars. 

Molecules more sensitive to UV irradiation are also potential 
tracers of the mechanically-heated layer, although they are 
trickier to employ. 
The abundances in the mechanically-heated layer of molecules 
that are dissociated 
by \Lya\ (Bergin et al.\ 2003) depend on the details of \Lya\ 
propagation through the FUV-heated layer. 
In addition, molecules like water are sensitive to 
mechanical heating in two ways, firstly because the 
abundance of water is sensitive to temperature 
(the synthesis of water through the neutral chemistry pathway  
weakens considerably below $\sim 250$\,K) 
and secondly through the excitation of the water 
transitions themselves. 
As a result, in the inner disk ($< 1$\,AU), the strength of water 
emission from the mechanically-heated layer 
depends on both \Lya\ propagation through the 
FUV-heated layer as well as mechanical heating. 
If we can better understand \Lya\ propagation, 
``lukewarm'' (200-500\,K) water emission,  
detectable at mid-infrared wavelengths, 
is a potentially 
useful diagnostic of a mechanically heated layer. 
At 0.5\,AU, the column density of lukewarm water is 
$\sim 1\times 10^{17}\psqcm$ in the reference model 
and more than 100 times smaller when mechanical heating 
is absent.

Our results expand on and endorse those of 
Hirose \& Turner (2011). They previously showed 
that while stellar irradiation dominates the heating 
in the surface layer of the disk at 1\,AU, 
turbulent dissipation of magnetic and kinetic energy 
can dominate the heating in the layer immediately 
below it (their Figure 1). 
By treating in detail the properties of the gaseous atmosphere, 
including the role of FUV irradiation, we are able to estimate 
the thermal-chemical properties of the 
layer in which mechanical heating may produce a 
distinctive observational signature. 

\subsection{The Heat Signature of Accretion in Context} 

Observational signatures of mechanical heating in inner disks 
($< 10$\,AU) 
would complement the ongoing search for evidence of the 
MRI in outer protoplanetary disks through the turbulent motions that they 
are expected to generate. 
Non-thermal line broadening,   
at the level of $\sim 50\%$ of the sound speed that is 
measured in spatially resolved line emission 
$\sim 200$\,AU away from the star,  
had been inferred and interpreted as evidence for turbulent line broadening 
generated by the MRI in outer disks 
(HD163296---Hughes et al.\ 2011; DM Tau---Guilloteau et al.\ 2012). 
More recent, higher resolution observations with ALMA of HD163296 
favor lower levels of non-thermal broadening, 
at the level of only $\lesssim 5$\% (Flaherty et al.\ 2015)  
or $\sim 10$\% (de Gregorio-Monsalvo et al.\ 2013) of the sound speed.  
Larger non-thermal line widths, 20--40\% of the sound speed,  
have been reported for the TW Hya disk 
beyond 40\,AU (Teague et al.\ 2016). 

In contrast to the outer disk, the inner, planet formation region 
of the disk is much less favorable to the MRI. 
The high disk column density at few AU distances 
($\sim 2000\gpsqcm$  at 1\,AU in the minimum mass solar nebula) 
shields the midplane from ionizing radiation, 
and the high midplane density enhances recombination. 
The resulting low ionization fraction deep in the disk produces a 
midplane ``dead zone'' in which 
the MRI is not expected to operate (Gammie 1996). 
As a result, any accretion is expected to be restricted 
to the disk surface.

Recent work on the impact of non-ideal MHD effects on the 
MRI suggest that ambipolar diffusion and the Hall effect 
suppress the MRI in the planet formation region 
(1--10\,AU) even in the surface layers of the disk; 
magneto-thermal disk winds, launched from the 
externally heated and ionized disk surface, have emerged 
as an alternative angular momentum removal mechanism at 
planet formation distances 
(Bai \& Stone 2013; Kunz \& Lesur 2013;
Gressel et al.\ 2015; Bai et al.\ 2016). 
The wind mass loss rates are thought to be large,  
a considerable fraction of the wind-driven disk accretion rate 
(Bai et al.\ 2016).  
Bai et al.\ (2016) suggest that 
observational evidence for such a massive wind might be found 
in the low-velocity blueshifted forbidden line emission from 
T Tauri stars (e.g., OI 6300\AA; Simon et al.\ 2016), 
although the thermal and dynamical properties of the wind are 
still uncertain enough that 
detailed predictions have not been made for comparison with 
observations.

Given the uncertain state of the theoretical predictions for 
angular momentum transport through disks or via disk winds, 
it seems prudent to look for ways to determine {\it observationally} 
whether the MRI is active or not at planet formation distances. 
One approach is to look for chemical evidence of turbulent 
motions 
produced by the MRI. If gas is transported (vertically or radially) in 
turbulent eddies quickly enough, it may be unable to equilibrate 
chemically to the local conditions (temperature, density, irradiation 
field), and a chemically distinctive signature of turbulent transport 
may be observable. 
This approach is challenging in inner disk 
atmospheres, because the chemical timescale is short or comparable to 
the dynamical time. 
In our models, the chemical timescale at 1\,AU is much shorter than 
1\,year for most species, similar to the results found by other 
investigators (e.g., Thi \& Bik 2005), 
which suggests a limited role for dynamical mixing in determining
the abundances of these species (cf.\ Heinzeller et al.\ 2011). 

Another possibility is to search for evidence of non-thermal 
line broadening, as in the millimeter studies, but on small 
(spatially unresolved) scales. 
Evidence for suprathermal line broadening at very small disk radii 
($< 0.3$\,AU) has been reported in systems with high disk accretion rates 
based on 2.3\micron\ CO overtone emission 
(e.g., Carr et al.\ 2004; Doppmann et al.\ 2009; Najita et al.\ 1996, 2009). 
These studies show that the closely-spaced lines 
near the $v$=2--0 CO bandhead can be used to probe 
the local line broadening, even with spatially unresolved data. 
When the lines at the bandhead are optically thick, 
the varying line overlap as a function of inter-line separation 
produced by local line broadening 
imprints a distinctive shape on the CO bandhead, 
and the local line broadening can be recovered 
with spectral synthesis modeling. 
Evidence for non-thermal line broadening from CO overtone emission 
is not surprising. 
The temperature of the gas that produces the CO overtone emission 
is high enough ($>1000$\,K) 
to thermally ionize alkali elements, and 
non-ideal MHD effects are expected to be weak,  
and the MRI active, as a result. 

Extending these measurements to the larger disk radii where 
a dead zone is expected (1--10\,AU)
requires diagnostics that probe cooler gas at larger radii
and also offer sets of closely spaced lines with a range of line spacings. 
Infrared transitions of water may be appropriate (Carr et al.\ 2004), as 
T Tauri stars commonly show a rich spectrum of water emission 
in the mid-infrared (Carr \& Najita 2008, 2011; Pontoppidan et al.\ 2010). 
This approach has yet to be attempted. 

Given the above challenges, searching for the ``heat signature'' of 
turbulent dissipation is a third, potentially promising option that 
we have explored here. 
To summarize,  
in our study of the effect of 
accretion related mechanical heating and 
energetic stellar irradiation (FUV and X-rays) 
on the thermal-chemical properties of disk atmospheres,  
we find that 
the effects of FUV heating, by both continuum and \Lya, 
are restricted to the upper molecular layer of the disk. 
The properties of the warm molecular layer, derived in the 
model, are generally in good agreement with the observed 
properties of disks inferred from spectroscopy at 
UV, NIR, and MIR wavelengths. 

With FUV heating and photodissociation restricted to the 
disk surface layer, 
the way is open for the MRI
to potentially signal its presence through 
accretion-related mechanical heating in the layer below. 
Warm CO, water, and other molecules are potential 
diagnostics of this region of the disk. 
To obtain robust predictions of signposts of mechanical heating, 
detailed work is needed on \Lya\ propagation through 
disk atmospheres and the chemistry of the region below the 
FUV-heated layer.

Some of our conclusions may be altered by the presence 
of magneto-thermal disk winds, which have been proposed as an 
alternative mechanism for disk angular momentum transport 
(Bai \& Stone 2016, 2013; 
Lesur et al.\ 2014; Gressel et al.\ 2015). 
If disk winds are massive, they can affect 
the predicted properties of disk atmospheres, e.g., by 
attenuating stellar FUV and X-rays before they reach the disk. 
A wind that is primarily atomic and entrains small dust 
grains as it leaves the disk atmosphere will 
scatter stellar \Lya\ photons and 
attenuate FUV photons. 
If considerably less \Lya\ and FUV continuum reaches the disk 
as a result, 
the good agreement between the properties of disk 
atmosphere and the molecular emission properties of disks 
described here would be compromised. The reduction in 
irradiation heating would increase the need for other heating 
sources such as mechanical heating. 
Future studies of irradiated atmospheres that include 
an intervening disk wind component would therefore lend 
insights into the properties of winds and the role of 
mechanical heating in disks. 

\acknowledgments
We thank the referee for valuable comments. 
This work was performed in part at the Aspen Center for Physics,
which is supported by National Science Foundation grant PHY-1066293.
It also benefitted from the National Science Foundation Grant 
No.\ NSF PHY11-25915 to the Kavli Institute for Theoretical Physics.
JN acknowledges the stimulating research environment supported by
NASA Agreement No.\ NNX15AD94G to the ``Earths in Other Solar Systems''
program. M\'A acknowledges support by NASA XRP grant NNX15AE24G. 
\newpage

\centerline{\bf Appendix A. Supplementary Figures}

The comparison in Figure 6 of the model properties of the 
warm molecular atmosphere with the molecular emission properties 
observed from disks requires calculations at radii 
beyond 0.5\,AU. In this Appendix, we present supporting model results 
at 1\,AU and 2\,AU (Figs.\ 7--10) 
that extend the results shown in Figures 1--4 to larger radii.  

Figure 7 shows the gas and dust temperatures, heating rates, 
and FUV irradiation fields at 0.5\,AU, 1\,AU, and 2\,AU in 
the reference model. 
While the 1/r$^2$ dilution of the stellar FUV and \Lya\ radiation 
field affects the peak temperature reached in the 
FUV-heated (strawberry-shaded) region, the general characteristics 
of the layered structure at 0.5\,AU (\S 3) are similar 
at the larger radii.  FUV photochemical heating still dominates 
in the warm upper molecular layer.  As the heating rates decrease
with disk radius and decreasing local volumetric density, the 
cooling rates (not shown) also decrease with radius, and a 
similar temperature structure is maintained out to 2\,AU
(see also ANG16 and AGN14).

Figure 8 shows the molecular abundances as a function of 
vertical column at 0.5\,AU, 1\,AU, and 2\,AU.  
The FUV-heated layer (strawberry shading) 
always has a significant, although not fully, molecular component 
as measured by $x(\hm )$.
The freeze-out of water beyond 1\,AU has a significant impact on
the molecular abundances; e.g., the gas phase abundances of 
water and OH are reduced in the mechanically-heated layer 
(chocolate shading). 
Two additional figures, which show the model results with and 
without \Lya\ irradiation 
(Fig.\ 9) and mechanical heating (Fig.\ 10), 
illustrate how the differences found at 0.5\,AU (\S 3) are 
qualitatively similar out to 2\,AU.

\newpage

\centerline{\bf Appendix B. Properties of 
Observed Molecular Diagnostics} 

As discussed in \S 3, 
the range of values shown in Figure 6 for the observed 
temperature, column density, and emitting area for each 
molecular diagnostic reflect the range of values measured 
from source to source and/or the uncertainty in the reported 
values. The observations underlying the molecular emission 
properties summarized in the figure are described in 
greater detail below. 
While the figure attempts to make a general comparison 
between the model results and the emission properties of CTTS 
as a group, more detailed comparisons could be made for 
individual sources, by comparing the warm columns reported for 
an individual source with the results of models tailored to its 
properties (e.g., its FUV continuum, \Lya, 
and X-ray luminosities, extent of grain settling, etc). 

Simple slab models 
with a single temperature and column density 
and a given emitting area are often employed 
to characterize the molecular emission. 
In general, the reported properties, derived assuming LTE, 
reproduce 
the observed spectra fairly well. The good agreement is 
surprising in some cases given the potential for non-LTE 
level populations 
(\water---Meijerink et al.\ 2009; 
HCN---Bruderer et al.\ 2015; 
\cotwo---Bosman et al.\ 2017); sub-thermal populations may 
bias the LTE fit to lower temperatures. 
Because real disks are expected to show variations in temperature 
and density as a function of disk radius and height,  
the fit parameters are average values over the disk emitting 
volume.  
The choice of features to include in a slab model also 
introduces systematic differences in the results 
between studies.

\noindent{\bf UV Fluorescent \hm } 

Populating the excited vibrational levels of \hm\ that can 
be fluoresced by \Lya\ photons requires temperatures 
$\sim 2500$\,K, if the excitation is thermal 
(Herczeg et al.\ 2004; Schindhelm et al.\ 2012). 
In their study of \hm\ emission from TW Hya, 
Herczeg et al.\ (2004) found an \hm\ excitation 
temperature of 2500\,K (+700\,K/-500\,K) and an 
\hm\ column density of 
$\log (N_{\hm}/\psqcm) = 18.5$ (+1.2/-0.8), and constrained the 
emission to arise from within 2\,AU of the star. 
Schindhelm et al.\ (2012) found similar emission properties 
for a larger number of T Tauri stars: 
\hm\ temperatures of $2500\pm 1000$\,K and 
$\log (N_{\hm}/\psqcm) = 19\pm 1$. 
The UV fluorescent \hm\ profiles reported by 
France et al.\ (2012) for non-transition T Tauri stars 
indicate that the emission arises from disk radii $\sim 0.1-1$\,AU. 
To capture these properties of the \hm\ emission, in Figure 6 we 
assume an \hm\ temperature range of 1500--3500\,K,  
an \hm\ column density of $\log (N_{\hm}/\psqcm) = 18-20$, 
and 
that the bulk of the emission arises from within 0.7\,AU.

\noindent{\bf CO Fundamental Emission:} 

Using LTE slab model fits to spectra of CO fundamental emission 
from classical T Tauri stars (CTTS), Najita et al.\ (2003) inferred 
that the emission is not optically thin, with  
typical temperatures 1100--1500\,K and 
line-of-sight CO column densities $\sim 10^{18}\psqcm$.
Studying a different set of CTTS, 
Salyk et al.\ (2011b, see also 2011a) 
inferred similar rotational excitation temperatures of 1000--1700\,K,  
line-of-sight CO column densities $10^{18}-10^{19} \psqcm,$
and projected emitting areas $\pi R_e^2$ 
of 0.18--0.7 AU$^2$ (their Table 5) 
corresponding to $R_e \simeq 0.25-0.5$\,AU. 
If we assume an average deprojection correction factor of 
$\sqrt 2$ for the emitting area, the equivalent 
$R_e \simeq 0.3-0.6$\,AU. 
While much of the emission may arise within 0.6\,AU, 
fitting the line profiles with Keplerian rotation models 
implies that the emission extends to larger radii, 
$>2$\,AU.

Similar conclusions were found by Banzatti \& Pontoppidan (2015) 
and Bast et al.\ (2011). 
Banzatti \& Pontoppidan (2015) decomposed the CO emission from 
CTTS into broad and narrow components. The shape of 
the broad component is fit as emission 
extending from $\sim 0.05-0.3$\,AU, and 
the narrow component as emission from $\sim 0.5-2$\,AU. 
In their study of the subset of CO emission sources with a single 
narrow peak and broad base ($\sim 15$\% of all sources), 
Bast et al.\ (2011) inferred 
lower rotational temperatures of $\sim 300-800$\,K; the lack of 
spatially extended emission constrained the origin of the 
emission to within a few AU of the central star.  
In Figure 6 we assume a CO temperature range (800--2000\,K) that  
emphasizes the warmer range of reported values, 
a vertical CO column density 
$\log (N_{\rm CO}/\psqcm) = 17.5-18.5$, 
and that the bulk of the emission arises from within 0.7\,AU. 

\noindent{\bf Spitzer Molecular Emission}

Spectra taken with the {\it Spitzer Space Telescope} have 
been modeled with simple LTE slab models to characterize 
the detected molecular emission. 
The slab models (which are parameterized by a temperature, 
column density and emitting area for each molecular species) 
reproduce the spectra fairly well. 
In their study of the IRS SH emission from a modest sample of 
classical T Tauri stars, Carr \& Najita (2011) reported 
best fitting water emission temperatures in a narrow range 
of values 575--650\,K, 
water column densities $4\times 10^{17}-2\times 10^{18}\psqcm$, and 
projected emitting areas $\pi R_e^2$ with $R_e=$0.8--1.5\,AU
based on fits to the emission at 12--16\micron. 
The HCN emission was found to be 
rotationally and vibrationally hot with 
temperatures 500--900\,K, 
best fitting HCN column densities 
$2\times 10^{16}-6\times 10^{16}\psqcm$  
corresponding to emission that is optically thick or 
marginally optically thick,  
and $R_e=$0.3--0.6\,AU. 
An optically thin fit is not excluded; if the 
HCN emission has an emitting area equal to that of the 
water emission from a given source,  the HCN emitting 
column density is 10 times smaller than in the optically thick 
fit, i.e., with a column density  
$2\times 10^{15}-6\times 10^{15}\psqcm$.

The properties of the \ctwohtwo\ and \cotwo\ emission are more weakly 
constrained. Assuming the \ctwohtwo\ emission has the same 
temperature and emitting area as HCN, the 
\ctwohtwo\ emitting column densities were found to be 
0.04--0.4 that of HCN. 
The best-fitting \cotwo\ temperature was lower than for HCN, in the 
range 200--600\,K assuming LTE, with the emission arising from 
an area corresponding to $R_e > 0.6-2$\,AU, 
and the \cotwo\ column density unconstrained. 
The low inferred \cotwo\ temperature may reflect an origin in cooler 
gas (farther out or deeper in the atmosphere) than HCN, 
or possibly non-LTE level populations 
(Bosman et al.\ 2017). 

Carr \& Najita (2011) did not fit the OH emission in the IRS SH 
module with an LTE slab model. 
The OH emission from classical T Tauri stars at IRS SH wavelengths 
is non-thermal in origin, with a characterisitic 
rotational temperature of $\sim 4000$\,K, and likely arises 
from hot OH produced by the photodissociation of water 
(Najita et al.\ 2010; Carr \& Najita 2011, 2014).  
In contrast, the lower rotational OH lines detected in the 
LH module probe rotationally cooler gas consistent with 
thermal emission. 
Carr \& Najita (2008) fit the LH OH emission from AA Tau, finding 
a temperature 575\,K, 
an OH column density $9 \times 10^{16}\psqcm$, 
and an emitting area corresponding to $R_e=$ 2.2\,AU. 

Complementing the above studies,  
Salyk et al.\ (2011a) modeled the Spitzer IRS spectra of 
a larger sample of stars with LTE slab models. 
To characterize the water emission, they fit 65 water features 
in the 10--35\micron\ range and inferred emission 
properties similar to 
those reported by Carr \& Najita (2011): 
projected water emitting areas corresponding to $R_e=0.8-2$\,AU, 
temperatures 400-800\,K, and 
water column densities in the range $\log (N_\water/\psqcm) \simeq 17.6-19$ 
with a typical column density of $\log (N_\water/\psqcm) = 18,$ 
where the ranges reflect the values found for individual CTTS. 

The other molecules were assumed to have the same 
emitting area as water. 
With this assumption, the HCN emission is optically thin, with 
column densities in the range 
$\log (N_{\rm HCN}/\psqcm)=15-16$
and temperatures 600--800\,K. 
The \ctwohtwo\ is found to be warmer, with 
a temperature $T \sim 1000$\,K
and a column density 0.1 of HCN. 
A similar column density ratio was found by Carr \& Najita (2011). 
The \cotwo\ emission allowed a range 
of temperatures 400--1000\,K
and $\log (N_\cotwo/\psqcm) = 14.5-16$. 
The OH doublets in the 17--30.5\um\ region were 
best fit with a warm temperature 900--1100\,K 
and column densities in the range $\log (N_{\rm OH}/\psqcm) \sim 15-16.3$.
The \ammonia\ column density was constrained to 
$\log (N_\ammonia/\psqcm) < 16$ 
for an assumed temperature of 400\,K.

To capture these properties of the {\it Spitzer} molecular 
emission, in Figure 6 we assume a temperature range of 
400--1000\,K  for water, HCN, \cotwo\, and \ammonia\ 
to span the range of reported values and their uncertainties. 
Because Salyk et al.\ (2011a) found higher temperatures for 
\ctwohtwo\ and OH, we assume 
a slightly larger temperature range of 400--1500\,K for \ctwohtwo\ 
and a higher temperature range of 800--2000K for OH. 
The range of observed water column densities is shown as an 
order of magnitude around the typical value of 
$\log (N_\water/\psqcm) = 18$. 
Typical column densities for HCN and \cotwo\ are shown as 
$\log (N/\psqcm)= 15-16$, with 
\ctwohtwo\ an order of magnitude lower. 
The OH column density is shown as 
$\log (N_{\rm OH}/\psqcm) = 15.3-16.5$. 
We also assume that the bulk of the emission from all of these 
species arises from 
within the typical water-emitting region, or $R_e= 1$\,AU.

\newpage

\noindent {\bf References}

\noindent \'Ad\'amkovics, M., Glassgold, A.\ E., \& Meijerink, R.\ 2011, ApJ, 736, 143

\noindent \'Ad\'amkovics, M., Glassgold, A.\ E., \& Najita, J.\ R.\ 2014, ApJ, 786, 135
(AGN14)

\noindent \'Ad\'amkovics, M., Najita, J.\ R.\ \& Glassgold, A.\ E.\ 2016, ApJ, 817, 82
(ANG16)

\noindent Akimkin, V., Zhukovska, S., Wiebe, D., et al.\ 2013, ApJ, 766, 8

\noindent Bai, X.\ \& Goodman, J.\ 2009, ApJ, 701, 737

\noindent Bai, X.\ \& Stone, J.\ M.\ 2013, ApJ, 769, 76

\noindent Bai, X., Ye, J., Goodman, J., \& Yuan, F.\ 2016, ApJ, 818, 152 

\noindent Balbus, S.\ A.\ \& Hawley, J.\ F.\ 1992, ApJ, 400, 610

\noindent Banzatti, A.\ \& Pontoppidan, K.\ M.\ 2015, ApJ, 809, 167

\noindent Bast, J.\ E., Brown, J.\ M., Herczeg, G.\ J., van Dishoeck, E.\ F., 
\& Pontoppidan, K.\ M.\ 2011, A\&A, 527

\noindent Bergin, E., Calvet, N., D'Alessio, P., \& Herczeg, G.\ J.\ 2003, ApJL, 591, L159

\noindent Bethell, T., \& Bergin, E.\ 2009, Sci, 326, 1675

\noindent Bethell, T.\ J., \& Bergin, E.\ A.\ 2011, ApJ, 739, 78

\noindent Bosman, A., Bruderer, S., \& van Dishoeck, E.\ F.\ 2017, arXiv:170108040

\noindent Bruderer, S., Harsono, D., \& van Dishoeck, E.\ F.\ 2015, A\&A, 575, 94

\noindent Carmona, A.\ 2010, Earth, Moon, \& Planets, 106, 71

\noindent Carr, J.\ S., Tokunaga, A.\ T., Najita, J., Shu, F.\ H., \& 
Glassgold, A.\ E.\ 1993, ApJ, 411, L37

\noindent Carr, J.\ S., Tokunaga, A.\ T., \& Najita, J.\ 2004, ApJ, 603, 213

\noindent Carr, J.\ S.\ \& Najita, J.\ R.\ 2008, Science, 319, 1504

\noindent Carr, J.\ S.\ \& Najita, J.\ R.\ 2011, ApJ, 733, 102

\noindent Carr, J.\ S.\ \& Najita, J.\ R.\ 2014, ApJ, 788, 66

\noindent Chiang, E.\ I.\ \& Goldreich, P.\ 1997, ApJ, 490, 368

\noindent D'Alessio, P., Calvet, N., Hartmann, L., Lizano, S., \& Cant\'o, J.\ 1999, 
ApJ, 527, 893

\noindent D'Alessio, P., Calvet, N., Hartmann, L., Franco-Hern\'andez, R., 
\& Serv\'in H.\ 2006, ApJ, 638, 314

\noindent D'Alessio, P., Calvet, N., Hartmann, L., Lizano, S., \& Cant\'o, J.\ 1999, ApJ, 527, 893

\noindent de Gregorio-Monsalvo, I., M\'enard, F., Dent, W., et al.\ 2013, AA, 557, 133

\noindent Doppmann, G.\ W., Najita, J.\ R., Carr, J.\ S., \& Graham, J.\ R.\ 2011, 
ApJ, 738, 112

\noindent Draine, B.\ T.\ \& Bertoldi, F.\ 1996, ApJ, 468, 269

\noindent Du, F., \& Bergin, E.\ A.\ 2014, ApJ, 792, 2

\noindent Ercolano, B., Clarke, C.\ J., \& Drake, J.\ J.\ 2009, ApJ, 699, 1639

\noindent Flaherty, K.\ M., Hughes, A.\ M., Rosenfeld, K.\ A., et al.\ 2015, 
ApJ, 813, 99

\noindent France, K., Schindhelm, E., Herczeg, G.\ J., et al.\ 2012, ApJ, 756, 171

\noindent Fraser, H.\ J., Collings, M.\ P., McCoustra, M.\ R.\ S., \& Williams, D.\ A.\ 2001, MNRAS, 327, 1165

\noindent Furlan, E., Calvet, N., D'Alessio, P., et al.\ 2005, ApJ, 628 65

\noindent Gammie, C.\ 1996, ApJ, 457, 355

\noindent Glassgold, A.\ E., Meijerink, R., \& Najita, J.\ R.\ 2009, ApJ, 701, 142

\noindent Glassgold, A.\ E., \& Najita, J.\ R.\ 2001, in ASP Conf.\ Ser.\ 244, Young Stars
Near Earth: Progress and Prospects, ed.\ R.\ Jayawardhana \& T.\ Greene (San
Francisco, CA: ASP), 251

\noindent Glassgold, A.\ E., Najita, J., \& Igea, J.\ 2004, ApJ, 615, 972 (GNI04)

\noindent Glassgold, A.\ E., \& Najita, J.\ R.\ 2015, ApJ, 810, 125

\noindent Gorti, U., \& Hollenbach, D.\ 2008, ApJ, 683, 287

\noindent Gressel, O., Turner, N.\ J., Nelson, R.\ P.\ \& McNally, C.\ P.\ 2015, 
ApJ, 801, 84

\noindent Guilloteau, S., Dutrey, A., Wakelam, V., et al.\ 2012, A\&A, 548, 70

\noindent Hartmann, L., Calvet, N., Gullgring, E., \& D'Alessio, P.\ 1998, ApJ, 495, 385

\noindent Heinzeller, D., Nomura, H., Walsh, C., \& Millar, T.\ J.\ 2011, ApJ, 731, 115

\noindent Herczeg, G.\ J., Wood, B.\ E., Linsky, J.\ L., Valenti, J.\ A., \& Johns-Krull, C.\ M.\ 2004, ApJ, 607, 369

\noindent Hirose, S.\ \& Turner, N.\ J.\ 2011, ApJ, 732, L30

\noindent Hughes, A.\ M., Wilner, D.\ J., Andrews, S.\ M., et al.\ 2011, ApJ, 727, 85

\noindent Jonkheid, B., Faas, F.\ G.\ A., van Zadelhoff, G.-J., \& 
van Dishoeck, E.\ F.\ 2004, AA, 428, 511

\noindent Kamp, I., \& Dullemond, C.\ P.\ 2004, ApJ, 615, 991

\noindent Kamp, I., \& van Zadelhoff, G.-J.\ 2001, A\&A, 373, 641

\noindent Kunz, M.\ W.\ \& Lesur, G.\ 2013, MNRAS, 434, 2295 

\noindent Lesur, G., Kunz, M.\ W., \& Fromang, S.\ 2014, A\&A, 566, 56

\noindent Li, X., Heays, A.\ N., Visser, R., et al.\ 2013, A\&A, 555, A14

\noindent Meeus, G., Salyk, C., Bruderer, S., Fedele, D., Maaskant, K., 
Evans, N.\ J.\ II, van Dishoeck, E.\ F.\ 2013, AA, 559, 84

\noindent Meijerink, R., Pontoppidan, K.\ M., Blake, G.\ A., Poelman, D.\ R., \& Dullemond, C.\ P.\ 2009, ApJ, 704, 1471

\noindent Najita, J., Carr, J.\ S., Glassgold, A.\ E., \& Valenti, J.\ A.\ 2007, in Protostars and Planets V, ed.\ B.\ Reipurth, D.\ Jewitt, \& K.\ Keil (Tucson: Univ. Arizona
Press), 507

\noindent Najita, J., Carr, J., S., Glassgold, A.\ E., Shu, F.\ H., 
\& Tokunaga, A.\ T.\ 1996, ApJ, 456, 292

\noindent Najita, J., Carr, J., S., \& Mathieu, R.\ D.\ 2003, ApJ, 589 931

\noindent Najita, J.\ R., \'Ad\'amkovics, M., \& Glassgold, A.\ E.\ 2011, ApJ, 743, 147

\noindent Najita, J.\ R., Doppmann, G.\ W., Carr, J.\ S., Graham, J.\ R., 
\& Eisner, J.\ A.\ 2009, ApJ, 691, 738

\noindent Najita, J.\ R., Carr, J.\ S., Strom, S.\ E., Watson, D.\ M., Pascucci, I.,
Hollenbach, D., Gorti, U., \& Keller, L.\ 2010, ApJ, 712, 274

\noindent Nomura, H., Aikawa, Y., Tsujimoto, M., Nakagawa, Y., \& Millar, T.\ J.\ 2007, ApJ, 661, 334

\noindent \"Oberg, K.\ I., Linnartz, H., Visser, R., \& van Dishoeck, E.\ F., 2009, 
ApJ, 693, 1209

\noindent Palla, F., Salpeter, E.\ E., Stahler, S.\ W.\ 1983, ApJ, 271, 632

\noindent Pontoppidan, K.\ M., Salyk, C., Blake, G.\ A., Meijerink, R., 
Carr, J.\ S., \& Najita, J.\ 2010, ApJ, 720, 887

\noindent Rab, C., Baldovin-Saavedra, C., Dionatos, O., Vorobyov, E., \& G\"udel, M.\ 
2016, SSR, 205, 3

\noindent Salyk, C., Glake, G.\ A., Boogert, A.\ C.\ A., \& Brown, J.\ M.\ 2009, 
ApJ, 699, 330

\noindent Salyk, C., Pontoppidan, K.\ M., Blake, G.\ A., Najita, J.\ R., \& Carr, J.\ S.\ 
2011a, ApJ, 731, 130

\noindent Salyk, C., Blake, G.\ A., Boogert, A.\ C.\ A., \& Brown, J.\ M.\ 2011b, ApJ, 743, 112 

\noindent Schindhelm, E., France, K., Herczeg, G.\ J., et al.\ 2012, ApJ, 756, L23

\noindent Simon, M., Pascucci, I., Edwards, S., et al.\ 2016, ApJ, 831, 169

\noindent Teague, R., Guilloteau, S., Semenov, D., et al.\ 2016, A\&A, 592, 49

\noindent Thi, W.-F.\ \& Bik, A.\ 2005, A\&A, 557, 570

\noindent Turner, N.\ J., Fromang, S., Gammie, C., et al.\ 2014, in Protostars and Planets VI, ed. H.\ Beuther et al.\ (Tucson, AZ: Univ.\ of Arizona Press), 411

\noindent Visser, R., van Dishoeck, E.\ F., \& Black, J.\ H.\ 2009, A\&A, 503, 323

\noindent Walsh, C., Nomura, H., Millar, T.\ J., \& Aikawa, Y.\ 2012, ApJ, 747, 114

\noindent Woitke, P., Kamp, I., \& Thi, W.-F.\ 2009, A\&A, 501, 383

\noindent Wolk, S.\ J., Harnden, F.\ R., Flaccomio, E., Micela, G., Favata, F., Shang, H., \& Feigelson, E.\ D.\ 2005, ApJSS, 160, 423 

\noindent Woods, P.\ M., \& Willacy, K.\ 2009, ApJ, 693, 1360

\noindent Yang, H., Herczeg, G.\ J., Linsky, J.\ L., et al.\ 2012, ApJ, 744, 121


\begin{figure}[t!]\begin{center}
\includegraphics[height=4.6in]{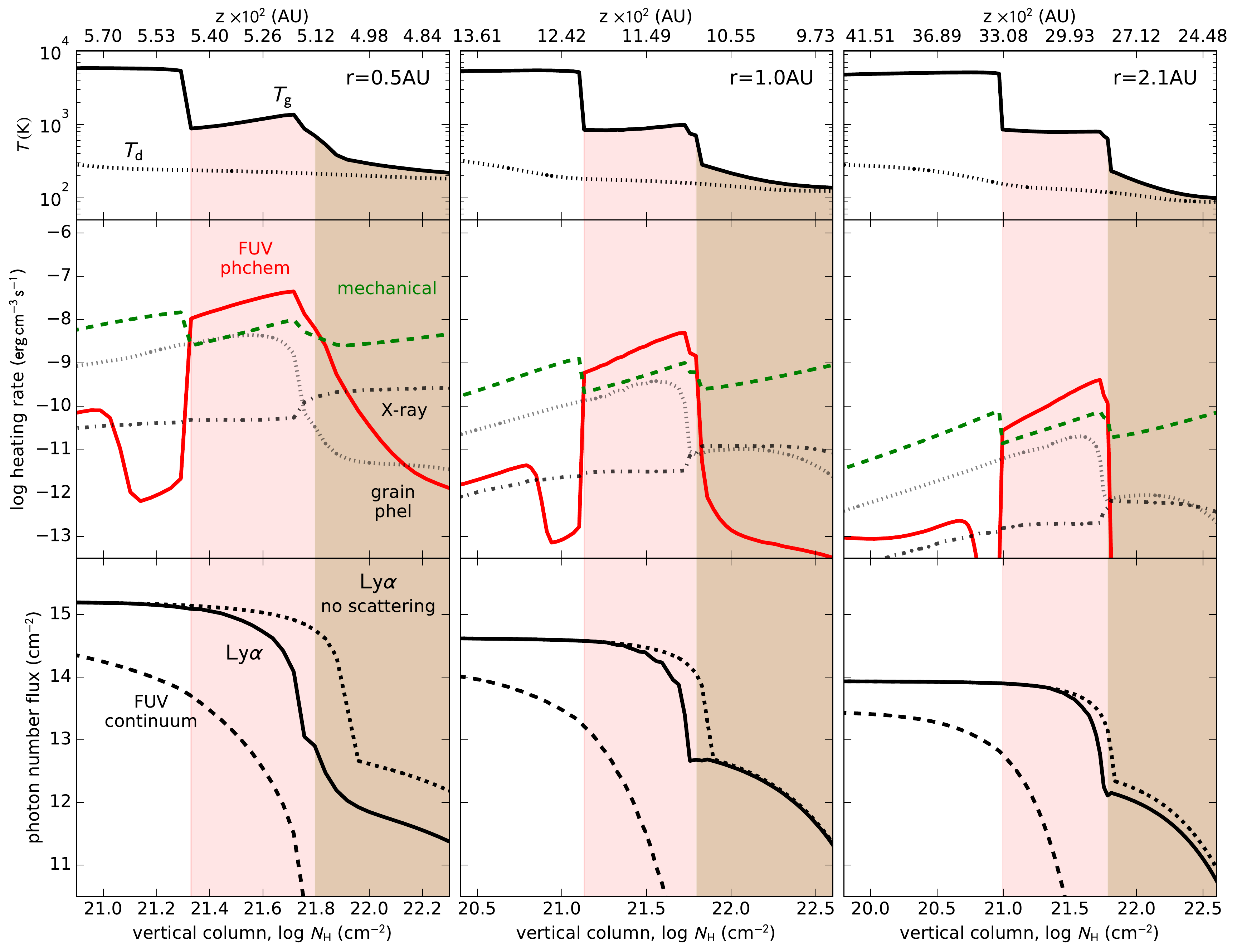}
\caption{Extension of Fig.~1 to 2\,AU. 
Vertical profiles of gas and dust temperatures (top row
of panels, solid and dotted lines, respectively), heating rates
(middle row) and FUV radiation fields (bottom row) at 
0.5\,AU, 1.0\,AU, and 2.1\,AU (panels in left, center and
right columns, respectively) in our reference model. FUV photochemical
heating (red solid line) dominates in the warm upper molecular layer
of the disk (strawberry shading) where the FUV continuum and \Lya\
are absorbed. Mechanical heating (green dashed line) dominates above
and below this region. X-ray (dash-dotted line) and grain photoelectric
heating (dotted line) are shown for comparison. The FUV continuum
and \Lya\ are strongly attenuated in the photochemically-heated layer.
}
\end{center} \end{figure}

\begin{figure}[t!]\begin{center}
\includegraphics[height=4.6in]{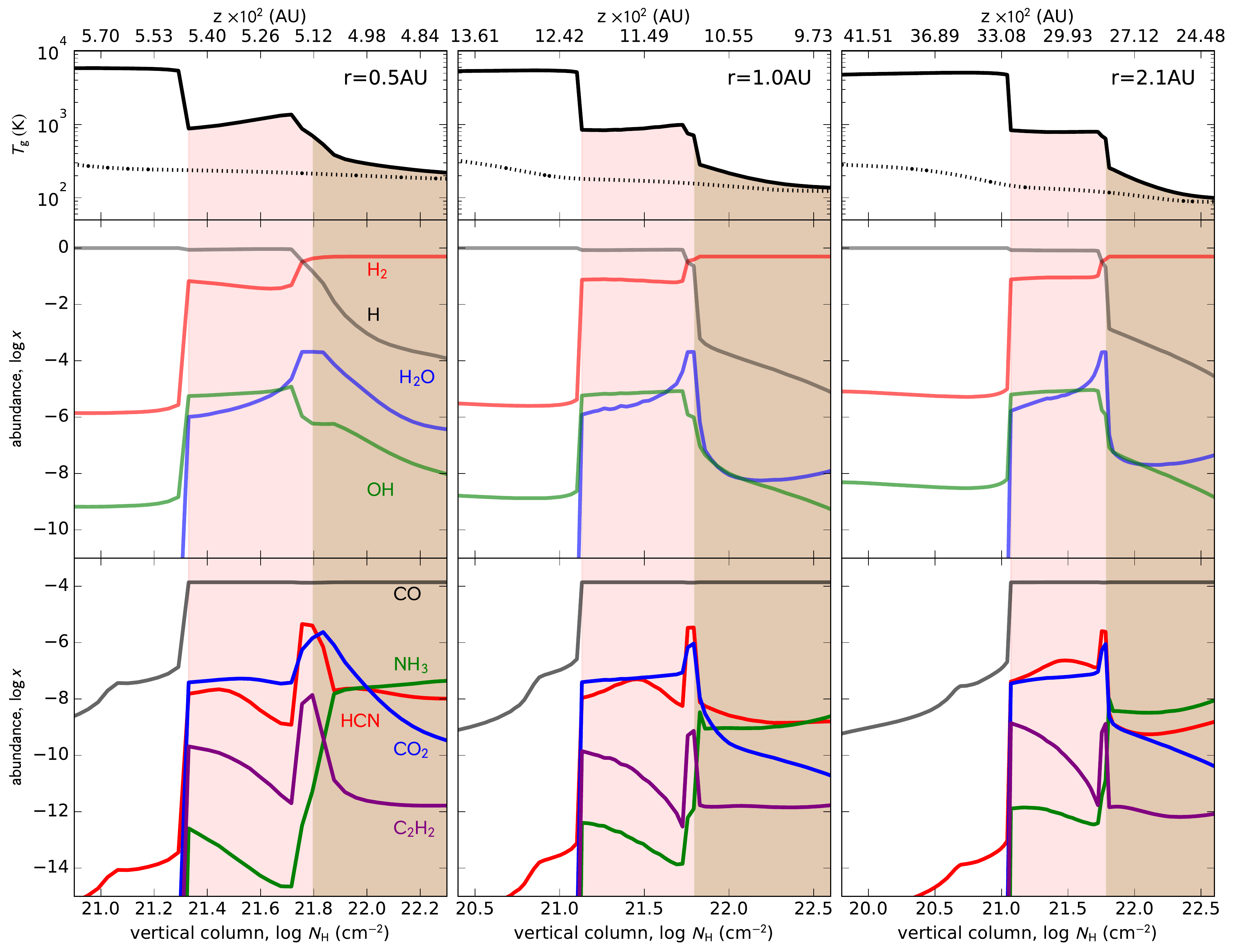}
\caption{Extension of Fig.~2 to 2\,AU. 
Vertical profiles of temperatures (top row of panels) and
abundances (middle and bottom row) at 0.5\,AU, 1.0\,AU, and
2.1\,AU (panels in left, center and right columns, respectively) for
our reference model. The warm upper layer where FUV heating dominates
(strawberry shading) is rich in molecular species that are commonly
detected from T Tauri disks. The cooler mechanically heated layer
(chocolate shading) lies below.
}
\end{center} \end{figure}

\begin{figure}[t!]\begin{center}
\includegraphics[height=3.1in]{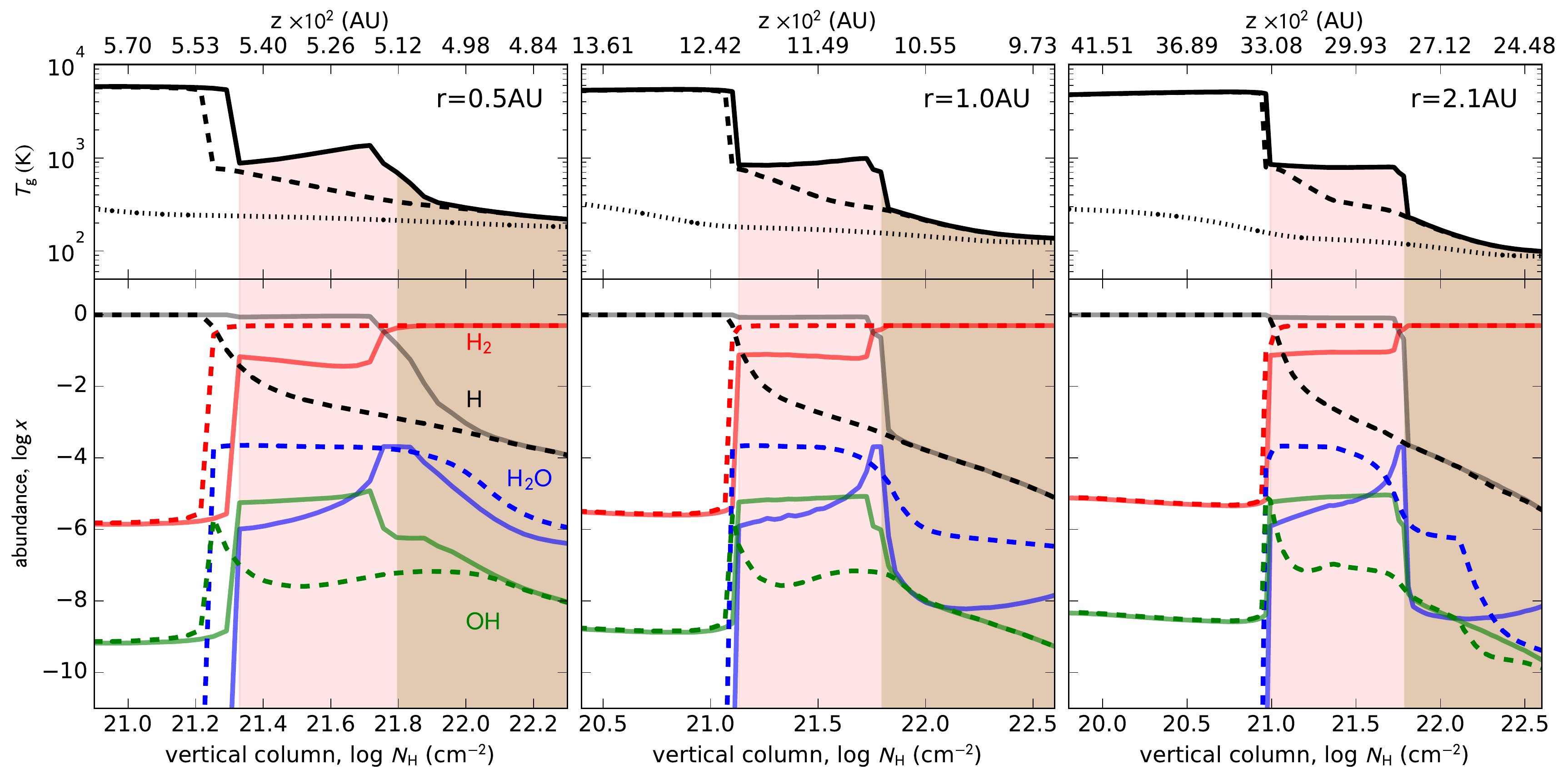}
\caption{Extension of Fig.~3 to 2\,AU. 
Vertical profiles of gas temperatures (top row of panels) and molecular
abundances (bottom panels) with and without \Lya\ irradiation (solid
and dashed lines respectively) at 0.5AU, 1.0, and 2.1AU (panels
in left, center and right columns, respectively). When \Lya\ is not
present, the molecular portion of the atmosphere is cooler, and the
molecular transition occurs higher in the atmosphere. When present,
\Lya\ dissociates \water\ and enhances the OH abundance in both the
FUV-heated (strawberry shading) and mechanically-heated (chocolate
shading) layers.
}
\end{center} \end{figure}

\begin{figure}[t!]\begin{center}
\includegraphics[height=3.1in]{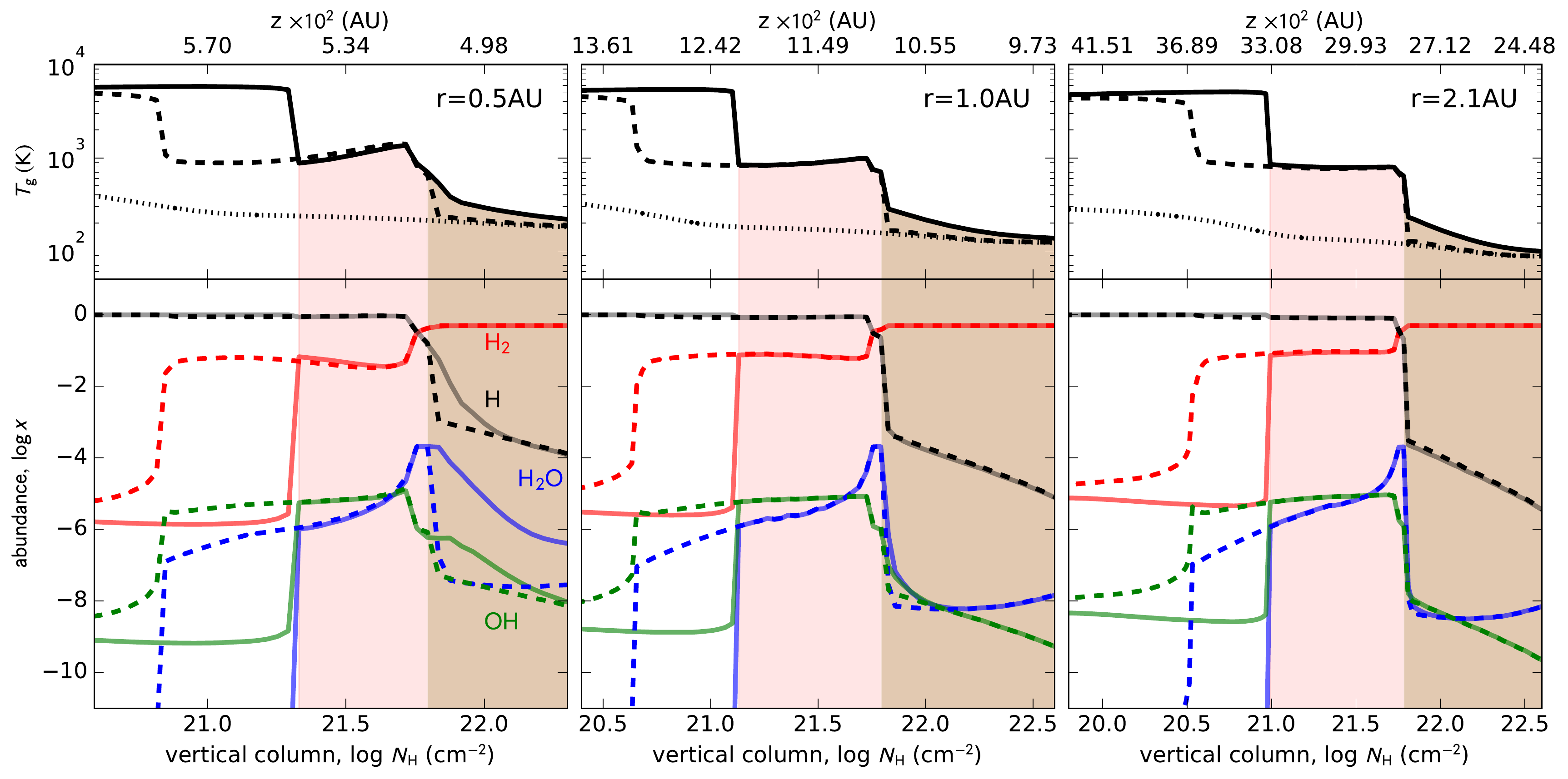}
\caption{Extension of Fig.~4 to 2\,AU. 
Vertical profiles of gas temperature (top row of panels) and 
molecular abundances (bottom panels) with and without mechanical heating 
(solid and dashed
lines, respectively) at 0.5\,AU, 1.0\,AU, and 2.1\,AU (panels in left,
center and right columns, respectively).  In models with reduced
mechanical heating, the molecular transition occurs higher in
the atmosphere and the temperature is reduced in the molecular
region below the FUV-heated layer.
}
\end{center} \end{figure}

\end{document}